\documentclass[aps,prb,reprint,floatfix,nobibnotes]{revtex4-1}
\pdfoutput=1
\usepackage{mathtools,amssymb}
\usepackage{units,siunitx}
\usepackage{graphicx}
\usepackage[version=3]{mhchem}

\begin{document}

\title{Semitransparent Polymer-Based Solar Cells with Aluminum-Doped Zinc Oxide Electrodes}

\author{Sebastian Wilken}
\email{sebastian.wilken@uol.de}
\author{Verena Wilkens}
\author{Dorothea Scheunemann}
\affiliation{Institute of Physics, Energy and Semiconductor Research Laboratory, Carl von Ossietzky University of Oldenburg, 26111 Oldenburg, Germany}
\author{Regina-Elisabeth Nowak}
\author{Karsten von Maydell}
\affiliation{NEXT ENERGY, EWE Research Centre for Energy Technology, Carl-von-Ossietzky-Str.~15, 26129 Oldenburg, Germany}
\author{J\"urgen Parisi}
\author{Holger Borchert}
\affiliation{Institute of Physics, Energy and Semiconductor Research Laboratory, Carl von Ossietzky University of Oldenburg, 26111 Oldenburg, Germany}

\begin{abstract}
With the usage of two transparent electrodes, organic solar cells are semitransparent and may be combined to parallel-connected multi-junction devices or used for innovative applications like power-generating windows. A challenging issue is the optimization of the electrodes, in order to combine high transparency with adequate electric properties. In the present work, we study the potential of sputter-deposited aluminum-doped zinc oxide~(AZO) as an alternative to the widely used but relatively expensive indium tin oxide~(ITO) as cathode material in semitransparent polymer--fullerene solar cells. Concerning the anode, we utilized an insulator\slash{}metal\slash{}insulator structure based on ultra-thin Au films embedded between two evaporated \ce{MoO3} layers, with the outer \ce{MoO3} film~(capping layer) serving as a light coupling layer. The performance of the ITO-free semitransparent solar cells is systematically studied as dependent on the thickness of the capping layer and the active layer, as well as the illumination direction. These variations are found to have strong impact on the obtained photocurrent. We performed optical simulations of the electric field distribution within the devices to analyze the origin of the current variations and provide deep insight in the device physics. With the conventional absorber materials studied herein, optimized ITO-free and semitransparent devices reached 2.0\% power conversion efficiency and a maximum optical transmission of 60\%, with the device concept being potentially transferable to other absorber materials.
\end{abstract}

\maketitle


\section{Introduction}
Recent developments in the field of organic photovoltaics~(OPV) led to promising power conversion efficiencies~(PCEs) around 7--9\% for single-junction devices on a laboratory scale.\cite{Liang2010,He2012} However, it is noteworthy that apart from these record devices most publications cover a significantly lower PCE level, in particular with regard to large-area organic solar cells.\cite{Dang2013,Jorgensen2013,Sondergaard2012} One major drawback of organic absorbers such as conjugated polymers, oligomers, or small molecules are the narrow absorption bands in comparison with their inorganic counterparts. Due to the incomplete match with the solar spectrum, achievable photocurrents of single-junction cells are comparatively low. A feasible approach to overcome this principle limitation is the development of multi-junction structures based on complementary absorber materials. The most commonly used approach is monolithic tandem\slash{}triple-junction cells with two\slash{}three subcells connected in series via suitable charge recombination layers.\cite{Ameri2009,Sista2011} The recent availability of high-performance low band gap polymers led to rapid progress of polymer-based tandem\slash{}triple-junction cells with efficiencies reaching up to 10.6\%.\cite{Li2013,You2013} According to Kirchhoff's circuit laws, photocurrent matching of the subcells is very critical for devices using a serial connection. This results in certain restrictions for the thickness of the absorber layers and, hence, may limit the overall performance of serial-connected devices.\cite{Dennler2007}

In general, a more ideal solution is the stacking of several semitransparent solar cells which are connected in parallel, so that subcells can be optimized separately in terms of maximum photocurrent generation.\cite{Sista2010} However, parallel-connected devices currently suffer from the lack of transparent electrodes with suitable electrical properties. Hence, the realization of efficient semitransparent OPV is of great interest and may also enable new fields of application like power-generating windows.\cite{Koeppe2009} The \textit{de facto} standard as transparent electrode in OPV is indium tin oxide~(ITO), which offers an excellent trade-off between conductivity and transmittance. However, concerning the limited abundance of indium, ITO is considered as bottleneck for the possible commercialization of the OPV technology.\cite{Emmott2012} Hence, the search for alternative transparent electrodes is an active research topic, with a variety of materials discussed in literature, including high-conductivity polymers,\cite{Gupta2013} ultra-thin metal films,\cite{Stec2012} metallic nanowires,\cite{Gaynor2010} graphene,\cite{Liu2013b} or carbon nanotubes.\cite{Hecht2011}

A further concept is the use of other transparent conductive oxides such as Al-doped ZnO~(AZO), which is commonly deposited by sputtering methods and widely established as cathode material in Cu(In,Ga)Se$_2$ and a-Si:H thin-film photovoltaics. However, relatively few publications report about the utilization of AZO electrodes in OPV devices.\cite{Murdoch2009,Sun2010,Park2010} Recently, Liu et al.\cite{Liu2013} achieved not only comparable results with AZO electrodes in polymer bulk heterojunction~(BHJ) solar cells compared to ITO, but also reported on suppressed photo-degeneration of the employed polymer--fullerene blend due to the UV-blocking capability of AZO. Regarding semitransparent BHJ solar cells, Bauer et al.\cite{Bauer2012} reported on the combination of ITO and AZO as bottom and top electrode, respectively, and reached a PCE of 4\% using a low band gap polymer--fullerene absorber blend layer.

Another approach are insulator\slash{}metal\slash{}insulator~(IMI) structures consisting of semitransparent metal films embedded between two dielectric layers,\cite{Cattin2013} which serve as charge transport and light coupling layer,\cite{OConnor2006} respectively. A variety of material combinations is reported for the realization of these multilayer electrodes, typically based on ultra-thin films of high-work function metals~(e.g., Au, Ag, Cu) and several metal oxides, including \ce{WO3}, \ce{MoO3}, \ce{V2O5}, \ce{NiO}, and \ce{ZnSnO}\slash{}\ce{ZnSnO3}. Concerning their application for organic solar cells, IMI structures are both implemented as bottom electrode in opaque\cite{Xue2014,Maniruzzaman2014,Sergeant2012,PerezLopez2012,Choi2011} and top electrode in semitransparent devices.\cite{Li2014,Shen2012,Tao2011,Winkler2011,Tao2009} Commonly, the IMI electrodes are applied via vacuum deposition techniques, which enables precise control of the optical and electrical properties by systematic variations of the particular layer thicknesses.

In this article, we report on the realization of ITO-free, semitransparent BHJ solar cells by combining AZO and IMI layers as bottom and top electrodes in an inverted device configuration. To demonstrate the applicability of this concept, BHJ blend layers consisting of poly(3-hexylthiophene)~(P3HT) and [6,6]-phenyl-C$_{61}$-butyric acid methyl ester~(PCBM) were employed. The AZO electrodes were grown via magnetron sputtering and additionally covered with an interfacial layer of (intrinsic) ZnO nanoparticles~(nc-ZnO), which is found to promote significantly improved electron collection properties. For the IMI electrodes, we employed an evaporated \ce{MoO3}\slash{}Au\slash{}\ce{MoO3} multilayer structure, based on ultra-thin Au films with a thickness of \unit[10--12]{nm}. The inner \ce{MoO3} layer was implemented as hole injection\slash{}extraction layer with a fixed thickness, whereas the thickness of the external \ce{MoO3} light coupling layer~(capping layer) was systematically varied.

It is well known that due to the thin layers used in OPV, thin film interference effects have a strong influence on the spatial distribution of the optical electric field intensity in the devices. For example, the insertion of an optical spacer between the photoactive layer and the reflective back electrode has been established as a powerful tool to optimize light absorption in conventional polymer--fullerene solar cells.\cite{Kim2006,Gilot2007} With regard to IMI electrodes, systematic variations of the dielectric layer thicknesses are widely applicated either to increase the electric field intensity within the active layer\cite{Chen2014,Sergeant2012} or manipulate the overall spectral transmission in case of semitransparent devices.\cite{Tao2011} A valuable tool in understanding and optimizing the optical properties of OPV devices are simulations based on the transfer-matrix method~(TMM).\cite{Pettersson1999,Peumans2003,Burkhard2010} Here, we provide deep insight into the photocurrent generation in our semitransparent solar cells as dependent on the thickness of the capping layer and the active layer as well as the illumination direction by the comparison of experimental data with optical simulations of the electric field distribution based on a TMM model.


\section{Experimental Methods}
\subsection{Materials} AZO films of varying thickness~(\unit[370--1270]{nm}) were fabricated by DC magnetron sputtering on precleaned glass substrates from a ceramic ZnO target with \unit[2]{wt.\%} \ce{Al2O3} content. The deposition was carried out under an \ce{O2}\slash{}Ar gas mixture with an \ce{O2}\slash{}Ar ratio of 0.5/100 at a pressure of \unit[$6 \cdot 10^{-3}$]{mbar} and a temperature of \unit[350]{$^\circ$C}. P3HT was purchased from Merck KGaA~(94.2\% regioregularity, Mw = \unit[54\,200]{g/mol}, Mn = \unit[23\,600]{g/mol}) and PCBM from Solenne BV~(99.5\% purity). 1:1~(wt:wt) mixtures of P3HT:PCBM were dissolved with varying total concentration in 1,2-dichlorobenzene~(Sigma-Aldrich, anhydrous, 99\%) and stirred at \unit[80]{$^\circ$C} over night. Colloidal ZnO nanoparticles were synthesized from zinc acetate dihydrate~(Sigma-Aldrich, 98\%) and potassium hydroxide~(Carl Roth, 85\%) according to previous reports.\cite{Pacholski2002,Wilken2012,Wilken2014} The as-obtained nanoparticles were dissolved in a 9:1~(v:v) mixture of chloroform and methanol with a concentration of \unit[15]{mg/ml}. ITO-coated glass substrates were purchased from Pr\"{a}zisions Glas \& Optik GmbH, Germany, with a sheet resistance of \unit[$\le 10$]{$\Omega$/sq}. \ce{MoO3} powder~(99.5\%) was supplied by Sigma-Aldrich.

\subsection{Material Characterization} UV-Vis transmission spectra were recorded with a spectrophotometer~(Varian Cary 100).  Sheet resistance was measured with a source measurement unit~(Keithley 2400) using the four-point probe method either in linear~(AZO, ITO) or van der Pauw configuration~(ultra-thin metal films). Further details regarding the resistivity measurements can be found in the Supplemental Material.\cite{SM} Electron microscopy was performed with a scanning electron microscope~(Fei Helios NanoLab 600i). Film thicknesses were monitored with a stylus profilometer~(Veeco Dektak 6M). Optical constants~($n$, $k$) of thin films, either on glass or Si, were determined by means of spectroscopic ellipsometry~(SE) using a rotating analyzer ellipsometer~(Woollam VASE). Experimental data was analyzed with a supplied software~(WVASE 32). Details regarding the SE measurements and the modeling of the experimental data can be found in the Supplemental Material.\cite{SM}

\subsection{Solar Cell Preparation} Glass substrates, coated either with AZO or ITO, were partially etched with hydrochloric acid, cleaned with water, detergent, acetone, and 2-propanol in an ultrasonic bath, and transferred in a \ce{N2} filled glove box. Subsequently, a \unit[$\sim\!\!40$]{nm} thick layer of ZnO nanoparticles was applied via spin-coating. Without further treatment of the nc-ZnO layer, the P3HT:PCBM blend was spin-coated at \unit[600]{rpm} from solutions with varying concentrations ranging from 10:10 to \unit[25:25]{mg/ml}, resulting in a layer thickness of 60, 100, 140, and \unit[200]{nm}, respectively. Afterwards, the samples were annealed on a hot plate at \unit[150]{$^\circ$C} for \unit[10]{min}. Finally, the transparent multilayer anode was thermally evaporated under high vacuum~(\unit[$10^{-6}$]{mbar}) in the sequence \ce{MoO3}~(\unit[12]{nm})\slash{}Au~(\unit[10-12]{nm})\slash{}\ce{MoO3}(\unit[$x$]{nm}) with $x = (0, 25, 50, 75)$. Film thicknesses were monitored \textit{in situ} using an individually calibrated quartz crystal microbalance. The photoactive area of the cells~(\unit[$\sim\!\!0.3$]{cm$^2$}) was delimited by the geometric overlap of the bottom~(AZO or ITO) and top~(\ce{MoO3}\slash{}Au\slash{}\ce{MoO3}) electrode and precisely measured for every individual cell with a stereoscopic microscope.

\subsection{Solar Cell Characterization} Current density--voltage~($J$--$V$) characteristics were re\-corded in ambient air with a semiconductor characterization system~(Keithley 4200). Four individual cells on each substrate were measured simultaneously. To study the photovoltaic performance, the samples were illuminated with a class AAA solar simulator~(Photo Emission Tech.), providing a simulated AM1.5G spectrum. The illumination intensity was adjusted to \unit[100]{mW/cm$^2$} using a calibrated Si reference cell. Spectral mismatch~($\le 5\%$) was not taken into account. External quantum efficiency spectra were measured with a custom-built system~(Bentham PVE300). The samples were illuminated by monochromatic light, chopped with a frequency of \unit[130]{Hz}, and the photocurrent was measured under short-circuit conditions using a lock-in amplifier~(Stanford Research Systems SR830).

\subsection{Optical Simulation} One-dimensional optical simulations of the semitransparent solar cells were performed using TMM modeling. Therefore, the devices were treated as a series of layers stacked on top of each other, and the electromagnetic field intensity at each position inside the layer stack was calculated based on a \texttt{MATLAB} program code developed by Burkhard and Hoke.\cite{Burkhard2010} The optical constants and the thickness of each layer served as input parameters for the simulation. The first and last layer of the stack, i.e., glass at the cathode side and air at the anode side, were treated as semi-infinite media. From the electromagnetic field distribution, optical device characteristics such as the active absorption within the P3HT:PCBM blend layer, parasitic absorption losses in the electrodes and interface layers, and the overall reflectance and transmittance of the solar cells were calculated. The exciton generation rate was estimated by convolution of the active absorption with the standard AM1.5G spectrum. Under the assumption of a certain wavelength-independent internal quantum efficiency, the photocurrent density was calculated by integrating the exciton generation rate over the thickness of the active layer.


\section{Results and Discussion}
\subsection{Overview over the Photovoltaic Performance}
\label{sec:overview}

\begin{figure*}
\centering 
\includegraphics{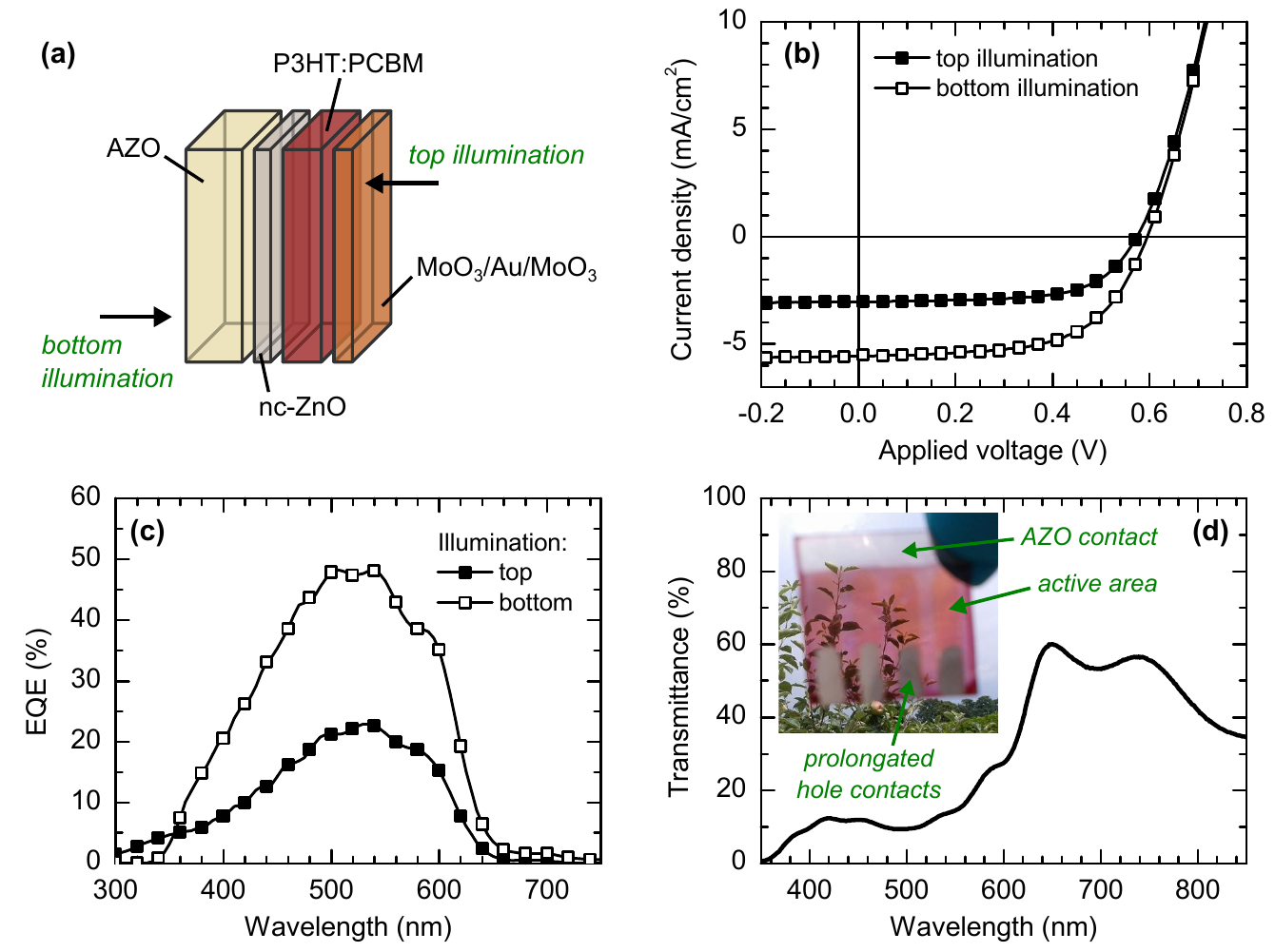}
\caption{(a)~Schematic device architecture of the inverted semitransparent BHJ solar cells. (b)~$J$--$V$ characteristics under simulated AM1.5G illumination~(\unit[100]{mW/cm$^2$}) and (c)~EQE spectra of an optimized device with the configuration AZO~(\unit[1270]{nm})\slash{}nc-ZnO~(\unit[40]{nm})\slash{}P3HT:PCBM~(\unit[140]{nm})\slash{}\ce{MoO3}~(\unit[12]{nm})\slash{}Au~(\unit[12]{nm})\slash{}\ce{MoO3}~(\unit[50]{nm}), excited via the \ce{MoO3}\slash{}Au\slash{}\ce{MoO3} multilayer anode~(top illumination) and the AZO cathode~(bottom illumination), respectively. (d)~UV/Vis transmittance through the whole device. The inset in panel~d shows a photograph of four individual cells on one substrate.}
\label{fig:fig1}
\end{figure*}

Figure~\ref{fig:fig1}a schematically presents the device architecture of the inverted semitransparent BHJ solar cells. To demonstrate the capability of the devices, Figure~\ref{fig:fig1}b shows $J$--$V$ curves of an optimized solar cell with the layer sequence AZO~(\unit[1270]{nm})\slash{}nc-ZnO~(\unit[40]{nm})\slash{}P3HT:PCBM~(\unit[140]{nm})\slash{}\ce{MoO3}~(\unit[12]{nm})\slash{}Au~(\unit[12]{nm})\slash{}\ce{MoO3}~(\unit[50]{nm}) with PCEs of 2.0\% and 1.1\% when illuminated through the AZO\slash{}nc-ZnO cathode~(called \textit{bottom illumination} in the following) and the \ce{MoO3}\slash{}Au\slash{}\ce{MoO3} multilayer anode~(called \textit{top illumination} in the following), respectively. The device exhibits pronounced rectification behavior  with an open-circuit voltage~($V_\text{oc}$) of \unit[0.59]{V}~(\unit[0.57]{V}) and a fill factor~(FF) of 0.60~(0.64) upon bottom~(top) illumination, which indicates that both the AZO\slash{}nc-ZnO and \ce{MoO3}\slash{}Au\slash{}\ce{MoO3} electrodes serve as appropriate interfaces for electron and hole collection, respectively, and ohmic losses, such as contact and sheet resistances, play a less important role. 

Consequently, the strong dependence of the PCE on the illumination direction is mainly related to the difference in the short-circuit current density~($J_\text{sc}$), which is \unit[5.6]{mA/cm$^2$} for bottom and \unit[3.0]{mA/cm$^2$} for top illumination. This finding is supported by the external quantum efficiency~(EQE) spectra shown in Figure~\ref{fig:fig1}c. Upon bottom illumination, the EQE reaches peak values of almost 50\% for wavelengths between 500 and \unit[550]{nm}, compared to only 23\% in the case of top illumination. However, upon bottom illumination, the EQE spectrum shows a strong decrease for wavelengths below \unit[400]{nm}, which is caused by parasitic absorption within the AZO layer and, hence, is not the case under illumination via the \ce{MoO3}\slash{}Au\slash{}\ce{MoO3} electrode. Therefore, both EQE spectra exhibit an intersection at a wavelength of about \unit[350]{nm}. Finally, Figure~\ref{fig:fig1}d visualizes the overall transparency of the device, reaching a transmittance of 35--60\% for wavelengths between 650 and \unit[900]{nm}.

At a first sight, the larger photocurrent upon illumination from the cathode side appears reasonable because of the higher far-field transmittance of AZO~(see Supplemental Material,\cite{SM} Figure~S1) compared to the \ce{MoO3}\slash{}Au\slash{}\ce{MoO3} multilayer system~(see Section~\ref{sec:var-anode}). However, at a closer look, the transparency of the electrodes alone cannot explain the nearly twofold increase in $J_\text{sc}$ upon bottom illumination. It is well-known that the electromagnetic field intensity within thin-film devices like organic solar cells is strongly governed by optical interference effects, caused by multiple reflections of the incident light at the layer interfaces. Hence, for a more detailed investigation, we modeled the optical electric field distribution within our semitransparent solar cells based on a TMM model.\cite{Pettersson1999,Peumans2003,Burkhard2010} As a prerequisite, the optical constants as a function of wavelength, i.e., the refractive index $n$ and the extinction coefficient $k$ as the real and imaginary part of the complex index of refraction, need to be known for all of the material layers. Therefore, we determined $n$ and $k$ for the involved materials by means of spectral ellipsometry~(SE), and the results as well as further details regarding the SE measurements and the modeling of the experimental data are given in the Supplemental Material~(Figure~S2).\cite{SM}

\begin{figure}
\centering 
\includegraphics{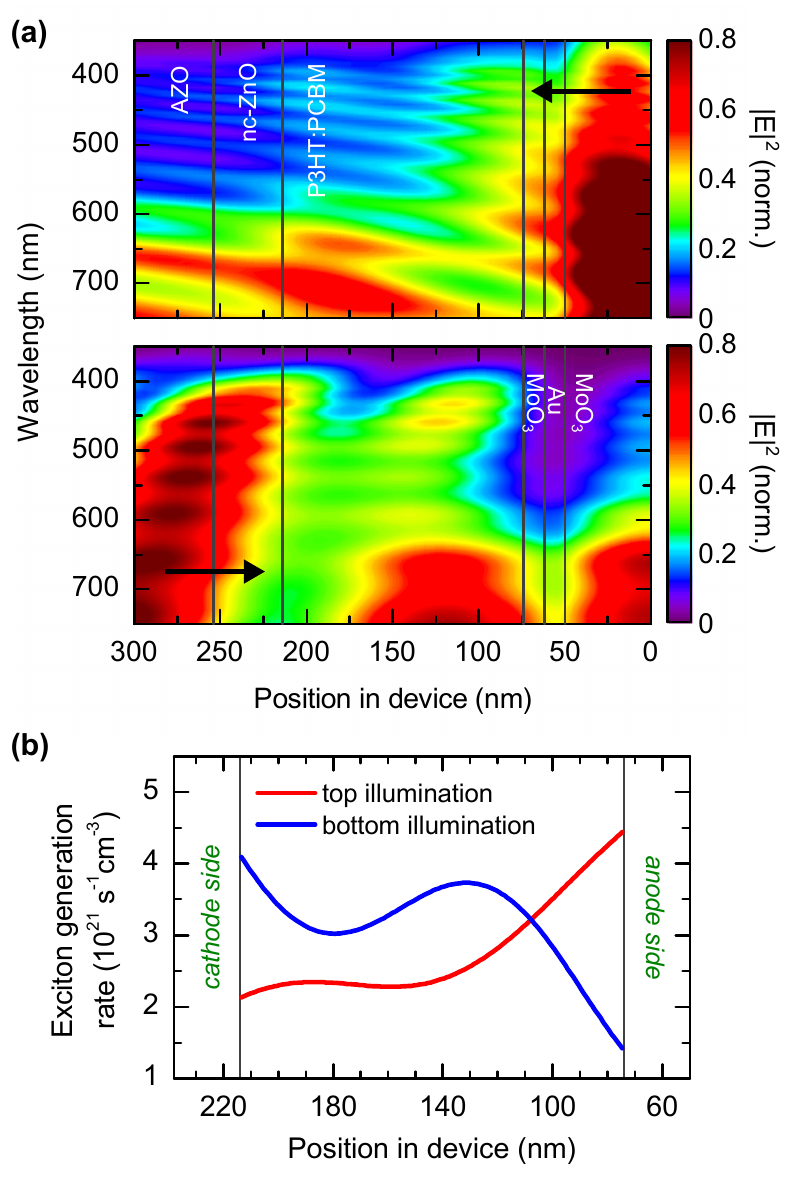}
\caption{(a)~Simulated electric field intensity $|E|^2$ for an inverted semitransparent BHJ solar cell in the optimized configuration AZO~(\unit[1270]{nm})\slash{}nc-ZnO~(\unit[40]{nm})\slash{}P3HT:PCBM~(\unit[140]{nm})\slash{}\ce{MoO3}~(\unit[12]{nm})\slash{}Au~(\unit[12]{nm})\slash{}\ce{MoO3}~(\unit[50]{nm}) upon top illumination~(upper panel) and bottom illumination~(lower panel). Arrows indicate the direction of incident light. (b)~Simulated exciton generation rate within the active layer upon AM1.5G illumination from the anode side~(top illumination) and the cathode side~(bottom illumination), respectively.}
\label{fig:fig2}
\end{figure}

In Figure~\ref{fig:fig2}a, the simulated spatial distribution of the electric field intensity $|E|^2$~(normalized to the incoming electric field intensity) is displayed for the optimized device configuration upon illumination from the anode and cathode side, respectively. It is obvious that $|E|^2$ strongly depends on the direction of incident light, particularly concerning the electric field intensity within the active layer, which is directly correlated to the amount of photogenerated excitons. Upon top illumination, the maximum field intensity is located close to the anode\slash{}absorber interface, followed by a steady decrease of $|E|^2$ along the optical path of incident light. This behavior qualitatively agrees with the Beer--Lambert law, which predicts an exponential attenuation of the electric field across absorbing media. In contrast, under illumination from the bottom side, another maximum of $|E|^2$ can be seen close to the center of the active layer for wavelength between 400 and \unit[500]{nm}, caused by internal reflection at the \ce{MoO3}\slash{}Au\slash{}\ce{MoO3} interface and subsequent interference with incident light.

The electric field intensity obtained from the TMM simulations serves as input for the calculation of device characteristics such as the active absorption, which is the fraction of incoming photons being absorbed throughout the active layer. Assuming that each absorbed photon is converted into one single exciton, the exciton generation rate can be estimated from the active absorption by convolution with the energy distribution of incident photons~(e.g., the AM1.5G solar spectrum). Using this procedure, we simulated the exciton generation rate profiles in the active layer for the optimized device configuration upon AM1.5G illumination, either from the anode or the cathode side~(Figure~\ref{fig:fig2}b). Corresponding to the spatial distribution of $|E|^2$, a high generation rate can be seen at the anode\slash{}absorber interface upon top illumination, whereas it is relatively low in the middle of the active layer and at the cathode interface. In contrast, a distinct peak appears close to the center of the absorber when the direction of incident light is changed to bottom illumination.

Integrating the exciton generation rate over the thickness of the active layer and making assumptions on the fate of the photogenerated excitons, one can also calculate the photocurrent density to be expected. In an ideal case, every generated exciton would be converted to extracted charge carriers. This corresponds to an internal quantum efficiency~(IQE) of unity and the photocurrent density expected in this case would be \unit[6.2]{mA/cm$^2$}~(top illumination) and \unit[7.1]{mA/cm$^2$}~(bottom illumination), respectively. It can clearly be seen that the experimentally obtained values of $J_\text{sc}$ are below these calculated values. In first place, this simply means that the IQE is not 100\%, which is not surprising for real solar cells. Several electrical loss mechanisms that are neglected in optical simulations can lead to IQE values below unity, including exciton recombination prior to charge separation and the (non-geminate) recombination of separated charge carriers during charge transport, either being mono- or bimolecular.

Nevertheless, under the assumption that the IQE is independent on wavelength, as it is usually observed for organic BHJ solar cells,\cite{Burkhard2009,Armin2014} the apparent IQE can be estimated by dividing the experimentally obtained $J_\text{sc}$ by the simulated photocurrent density.\cite{Slooff2007} For our optimized devices, this results in IQE values of $\sim\!\!50\%$~(top illumination) and $\sim\!\!80\%$~(bottom illumination). Hence, we can conclude that the nearly twofold increased photocurrent observed upon bottom illumination mainly arises from the higher IQE obtained in this case, the latter apparently being a direct consequence of the different exciton generation rate profile throughout the active layer upon illumination from the anode and cathode side, respectively.

To gain deeper understanding of the photocurrent generation in our semitransparent solar cells, we report in the following sections on systematic variations of the bottom electrode~(Section~\ref{sec:var-cathode}), the top electrode~(Section~\ref{sec:var-anode}), and the active layer~(Section~\ref{sec:var-thickness}).

\subsection{Comparison between AZO and ITO as Electron Contact}
\label{sec:var-cathode}

\begin{figure}
	\centering 
	\includegraphics{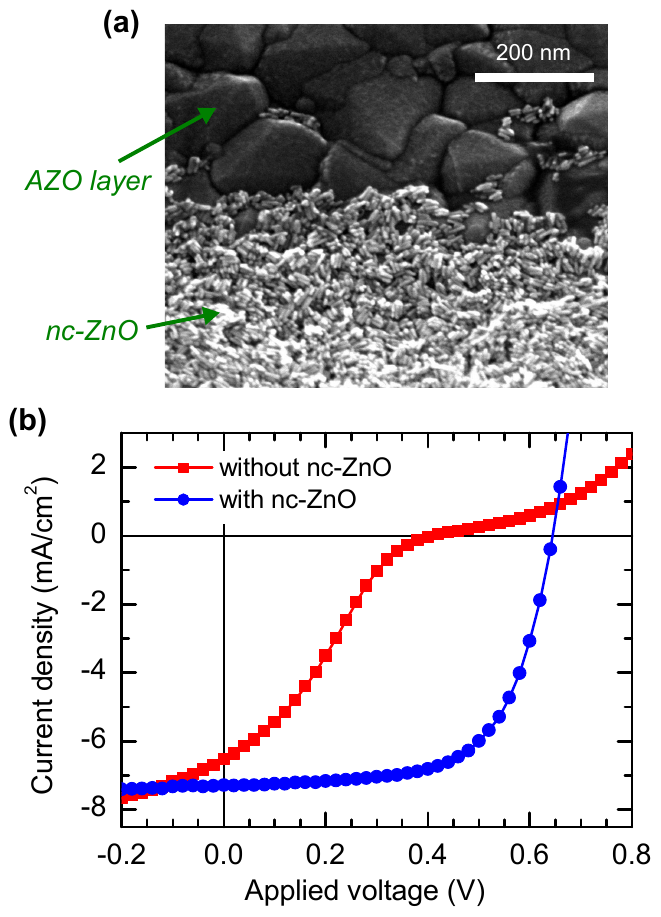}
	\caption{(a)~SEM image of an AZO electrode coated with a layer of ZnO nanoparticles~(nc-ZnO). An area at the border is selected here to visualize also the AZO below the nc-ZnO. (b)~Typical $J$--$V$ curves under simulated AM1.5G illumination for non-transparent solar cells~(anode: \ce{MoO3}~(\unit[12]{nm})\slash{}Ag~(\unit[100]{nm})) with and without a nc-ZnO interlayer between the AZO electrode and the active layer.}
	\label{fig:fig3}
\end{figure}

As an alternative to ITO, cathodes based on AZO were developed in this work. The AZO films were prepared via magnetron sputtering on glass substrates. Figure~S1 in the Supplemental Material\cite{SM} shows the spectral transmittance and sheet resistance of AZO layers with different thicknesses between \unit[370]{nm} and \unit[1270]{nm}. Both the average transmittance in the visible range and the sheet resistance were found to decrease with increasing film thickness. Figure~\ref{fig:fig3}a shows a SEM image of an AZO electrode, partially covered with a \unit[$\sim\!\!40$]{nm} thick interfacial layer of nanocrystalline ZnO~(nc-ZnO) prepared by spin-coating. For the latter, colloidally synthesized ZnO particles with an elongated shape and an average crystallite size of \unit[$27 \times 8$]{nm} were used, which are described in more detail elsewhere.\cite{Wilken2012,Wilken2014} It can be seen that the bare AZO electrodes had a relatively rough surface texture with a grain size of \unit[$\sim\!\!120$]{nm}, which is 	considerably smoothed by the application of the nc-ZnO layer.

To study the general applicability of the AZO electrodes, we first investigated non-transparent inverted solar cells with an opaque \ce{MoO3}~(\unit[12]{nm})\slash{}Ag~(\unit[100]{nm}) top contact. Figure~\ref{fig:fig3}b shows $J$--$V$ curves of exemplary devices, processed either on bare AZO electrodes without any surface modification~(red squares) or AZO electrodes modified with nc-ZnO~(blue circles). For the device without an additional nc-ZnO layer, the $J$--$V$ curve exhibits a pronounced S-shape behavior, resulting in poor photovoltaic performance with a $V_\text{oc}$ of \unit[0.41]{V}, a $J_\text{sc}$ of \unit[6.5]{mA/cm$^2$}, a FF  of 0.27, and a PCE of 0.7\%. In contrast, the S-shape is found to disappear and good rectifying behavior with improved values of $V_\text{oc}$~(\unit[0.64]{V}), $J_\text{sc}$~(\unit[7.3]{mA/cm$^2$}), and FF~(0.64) was obtained for the device with an additional nc-ZnO layer introduced between the AZO electrode and the active layer, leading to a more than fourfold increased PCE of 3.0\%. The latter is a typical value for the P3HT:PCBM system\cite{Dang2011} and indicates that the modified AZO\slash{}nc-ZnO electrode is well functioning in terms of electron injection\slash{}extraction and transport.

S-shaped $J$--$V$ curves are often observed for organic solar cells and attributed to non-ideal electrical properties of contact or interface layers.\cite{Liu2013,Reinhard2011,Schmidt2009,Kumar2009} One common explanation for such a behavior is the presence of energetic barriers for the injection and/or extraction of charge carriers at~(at least) one of the electrodes, which can effectively be described in terms of a reduced surface recombination velocity for majority charge carriers.\cite{Wagenpfahl2010,Sandberg2014} Note that an ideal contact would have an infinite surface recombination velocity, which means that all photogenerated charges within the active layer can be extracted. However, when the surface recombination velocity is reduced, majority charge carriers may accumulate at the vicinity of the electrode, resulting in the formation of space charge and, finally, an S-shaped $J$--$V$ curve. Hence, the S-shaped $J$--$V$ characteristics observed without an additional nc-ZnO layer suggest that the bare AZO electrodes without any surface modification posses poor electron extraction properties. Furthermore, the reduced values of $V_\text{oc}$ indicate a significant energy loss at the cathode when no additional nc-ZnO layer is applied. This points to undesired recombination of photogenerated holes at the cathode, probably related to the high doping level of AZO. Instead, when the additional nc-ZnO layer is applied, which is of intrinsic nature, the recombination is reduced due to the hole blocking property of intrinsic ZnO. A similar effect has been reported by Ihn et al.,\cite{Ihn2011} who investigated Ga-doped ZnO electrodes with and without an additional layer of intrinsic ZnO prepared by radio-frequency sputtering.

\begin{figure}
	\centering 
	\includegraphics{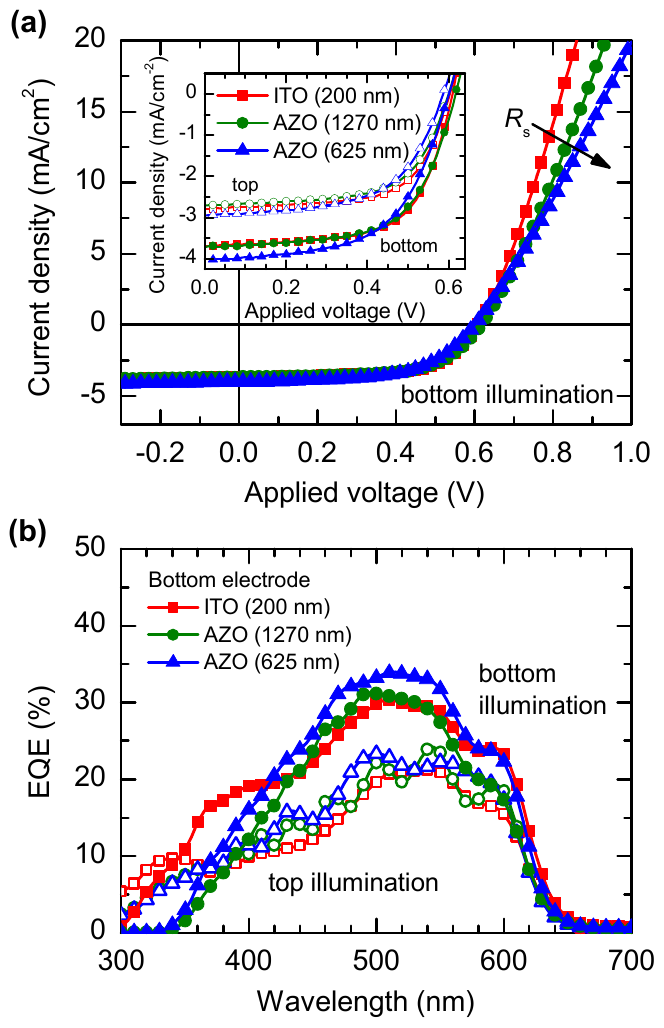}
	\caption{(a)~Averaged $J$--$V$ curves and (b)~representative EQE spectra for semitransparent solar cells with different bottom electrode materials~(ITO vs. AZO) and the anode system \ce{MoO3}~(\unit[12]{nm})\slash{}Au~(\unit[10]{nm})\slash{}\ce{MoO3}~(\unit[50]{nm}) upon top and bottom illumination. The arrow in panel~a indicates the increasing series resistance~($R_\text{s}$).}
	\label{fig:fig4}
\end{figure}

\begin{table*}
\caption{Photovoltaic parameters for semitransparent solar cells with different cathode materials~(ITO vs. AZO) upon top and bottom illumination.\protect\footnote{Anode: \ce{MoO3}~(\unit[12]{nm})\slash{}Au~(\unit[10]{nm})\slash{}\ce{MoO3}~(\unit[50]{nm}). The values reported are mean values averaged over 6-7 individual cells and the corresponding standard deviation.}}
\label{tab:tab1}
\centering
\begin{ruledtabular}
\begin{tabular}{ccccccc}
Cathode & Illumination &  $V_\text{oc}$~(V) & $J_\text{sc}$~(mA/cm$^2$) & FF & PCE~(\%) & $R_\text{s}$~($\Omega$cm$^2$)\\
\noalign{\vskip .75mm}
\hline
\noalign{\vskip .75mm}  
ITO (\unit[200]{nm}) & top & 0.60 $\pm$ 0.01 & 2.8	$\pm$ 0.2 & 0.64 $\pm$ 0.03 & 1.1	$\pm$ 0.1 & 9	$\pm$ 2 \\
AZO (\unit[625]{nm}) & top & 0.59 $\pm$ 0.01 & 3.0 $\pm$	0.2 & 0.59 $\pm$ 0.04 & 1.0	$\pm$ 0.1 & 15 $\pm$ 1 \\
AZO (\unit[1270]{nm}) & top & 0.60 $\pm$ 0.03 & 2.7	$\pm$ 0.3 & 0.63 $\pm$ 0.03 & 1.0	$\pm$ 0.2 & 11	$\pm$ 4   \\[.5em]
ITO (\unit[200]{nm}) & bottom & 0.61 $\pm$ 0.01 &	3.7	$\pm$ 0.3 &	0.64 $\pm$ 0.02 &	1.5	$\pm$ 0.2 &	8	$\pm$ 1 \\
AZO (\unit[625]{nm}) & bottom & 0.60 $\pm$ 0.01 &	4.0	$\pm$ 0.3 &	0.58 $\pm$ 0.03 &	1.4	$\pm$ 0.1 &	15	$\pm$ 2 \\
AZO (\unit[1270]{nm}) & bottom & 0.62 $\pm$ 0.02 &	3.7	$\pm$ 0.4 &	0.62 $\pm$ 0.03 &	1.4	$\pm$ 0.2 &	12	$\pm$ 4 \\
\end{tabular}
\end{ruledtabular}
\end{table*}

To investigate, whether the AZO\slash{}nc-ZnO system can compete as transparent electrode with the established material ITO~(here more precisely ITO\slash{}nc-ZnO), both electrodes were compared in one series of solar cells. Two thicknesses of the AZO layer were tested in these experiments, \unit[625]{nm} and \unit[1270]{nm}. Figure~\ref{fig:fig4}a shows $J$--$V$ curves for the different configurations, averaged over several individual cells, and Table~\ref{tab:tab1} summarizes the photovoltaic parameters. The $J$--$V$ curves show a pronounced difference in the photocurrent depending on the illumination direction for the reasons outlined in the previous section. 

However, we note that the contrast in $J_\text{sc}$ between top and bottom illumination is less pronounced than for the optimized devices because thinner Au films with a nominal thickness of \unit[$\sim\!\!10$]{nm} were used at this stage of the work. It is well known that the optical and electrical properties of ultra-thin metal films are very sensitive to small variations of the thickness near the percolation threshold, which is \unit[6]{nm} for Au in an ideal case.\cite{Hovel2010} Here, the thinner films exhibited a lower reflectivity and, hence, a smaller fraction of incident light was internally reflected at the anode\slash{}absorber interface compared to thicker Au films, leading to lower values of the photocurrent upon bottom illumination.

In contrast, the photocurrent was found to depend only slightly on the nature of the cathode and we obtained very similar values of the PCE for all three configurations under consideration. The small enhancement of $J_\text{sc}$ in case of the thin AZO electrodes~(\unit[625]{nm}) may be related to their higher average transmittance~(see Supplemental Material,\cite{SM} Figure~S1), however, our data provide no statistical evidence for this effect. More significant is the variation of the series resistance~($R_\text{s}$). Details regarding the determination of the values for $R_\text{s}$ given in Table~\ref{tab:tab1} can be found in the Supplemental Material~(Figure~S3).\cite{SM} The fact that the series resistance was found to be nearly independent of the illumination direction suggests that it is mainly determined by the sheet resistance of the cathode materials. The lowest values for $R_\text{s}$ were found for the ITO electrodes, followed by the thick~(\unit[1270]{nm}) and thin~(\unit[625]{nm}) AZO layers, which also had slight influence on the fill factor of the devices. The higher series resistance in case of the thin AZO layers can be explained by their larger sheet resistance~(\unit[14]{$\Omega$/sq}) compared to the thick ones~(\unit[6]{$\Omega$/sq}). The EQE measurements~(Figure~\ref{fig:fig4}b) support the finding that the photocurrent is more or less the same for ITO and AZO. In the range where the P3HT component of the active layer has a strong absorption, the devices with ITO electrodes provided a lower quantum efficiency than those with AZO layers, in particular in the case of bottom illumination. However, this is compensated by the wavelength region below \unit[400]{nm}, where the EQE is higher in the case of ITO.

\begin{figure*}
	\centering 
	\includegraphics{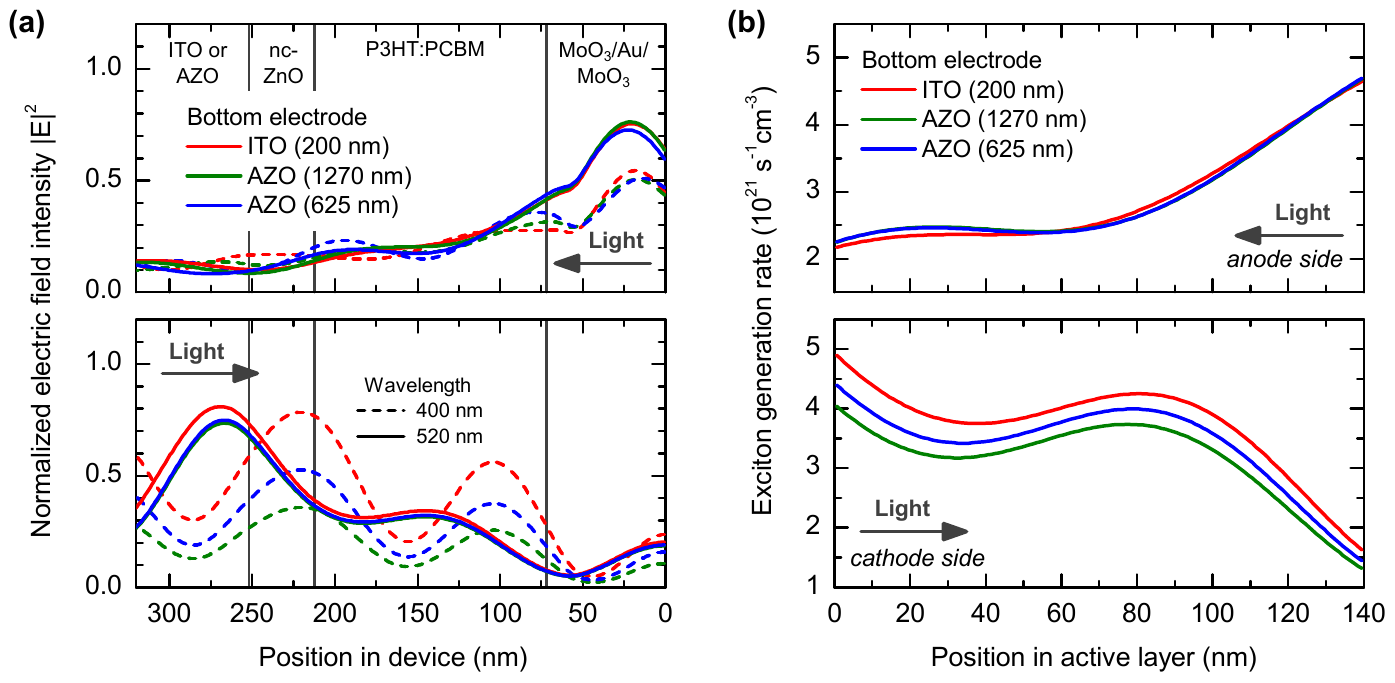}
	\caption{Results of the optical simulations for different bottom electrode materials~(ITO vs. AZO). (a)~Spatial distribution of the normalized electric field intensity for incident light with a wavelength of \unit[400]{nm} and \unit[520]{nm}, respectively, either from the anode side~(upper panel) or the cathode side~(lower panel). (b)~Resulting exciton generation rate profiles for AM1.5G illumination.}
	\label{fig:fig5}
\end{figure*}

To corroborate the observations made in the electric measurements, we simulated the spatial distribution of the electric field intensity in the devices. Figure~\ref{fig:fig5}a shows exemplarily the results for two selected wavelengths. At \unit[520]{nm}, which corresponds to the region around the maximum absorption of P3HT, the field intensity distribution inside the active layer does not significantly depend on the cathode configuration, this statement being valid for both illumination directions. However, there are also wavelengths which can be absorbed and where the distribution is not independent of the electrode. This is exemplarily shown for a wavelength of \unit[400]{nm}, where significantly different intensities can be seen, in particular in the case of bottom illumination, which is in reasonable agreement with the EQE measurements~(Figure~\ref{fig:fig4}b). To conclude on the photocurrent density to be expected, one has to consider all wavelengths which can be absorbed in the active layer. Hence, we calculated the spatially resolved exciton generation rate within the active layer, and the results are shown in Figure~\ref{fig:fig5}b. It can be seen that the exciton generation rate is nearly independent of the different cathodes tested in the case of top illumination, and shows only little variation in the case of bottom illumination. Particularly, no shifting of the peak positions within the active layer can be observed for the different cathode materials. This is consistent with the experimental finding that the photocurrent density depends strongly on the illumination direction, but barely on the usage of ITO or AZO. Thus, to conclude on this section, sputtered AZO with sufficiently small sheet resistance was found to be a suitable replacement for the the established material ITO.

\subsection{Variation of the MoO$_3$\slash{}Au\slash{}MoO$_3$ Multilayer Anode}
\label{sec:var-anode}

It is well known that the conductivity and transparency of ultra-thin metal films are highly sensitive to the film thickness.\cite{OConnor2008,Hovel2010} Hence, as a preliminary stage of the IMI electrodes, we investigated the sheet resistance, visible transmittance and morphology of Au films with variable thickness, evaporated on \ce{MoO3}~(\unit[12]{nm}) covered glass substrates~(see Supplemental Material,\cite{SM} Figures S4 and S5). For the examined Au thicknesses~(\unit[8--15]{nm}), a straight relation between average transmittance and sheet resistance was found, with the latter ranging from 4 to \unit[9]{$\Omega$/sq}, being lower, or at least comparable to the AZO and ITO electrodes used in this study. These values are significantly smaller than observed in an earlier study where the Au films were directly evaporated on glass,\cite{Wilken2012b} which indicates the importance of the \ce{MoO3} seed layer for film quality issues, e.g., surface coverage, roughness, and grain size.\cite{OConnor2008} The high film quality could also be observed by SEM. Additionally, Ag was tested as an alternative for Au, but yielded higher sheet resistance due to island growth~(see Supplemental Material,\cite{SM} Figures S4 and S5).

\begin{figure}
	\centering 
	\includegraphics{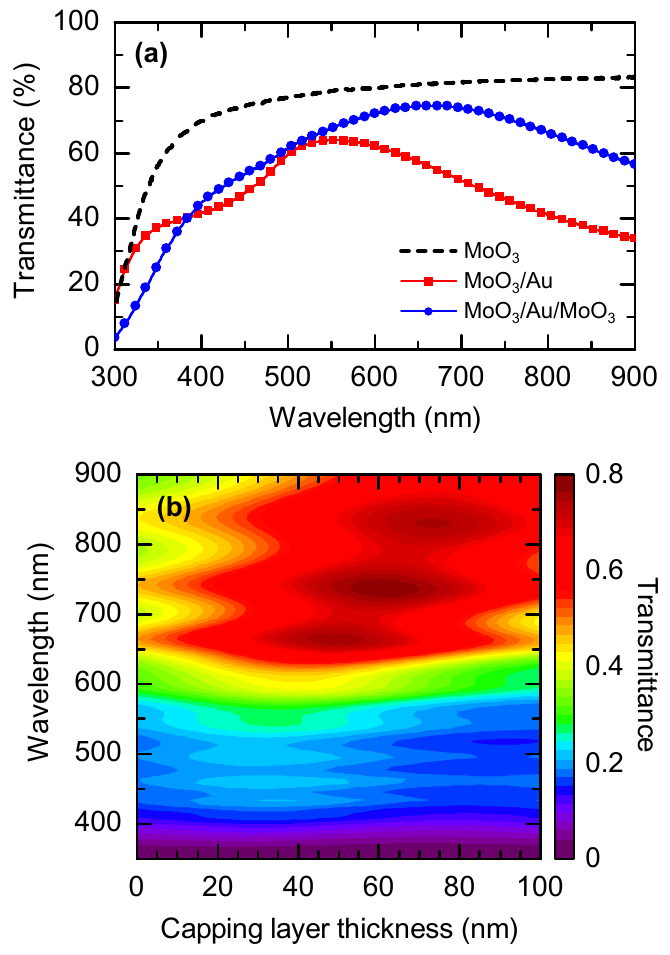}
	\caption{(a)~Measured transmittance through glass slides coated with a single layer of \ce{MoO3}~(\unit[12]{nm}) as well as the anode system with additional capping layer~(\ce{MoO3}~(\unit[12]{nm})\slash{}Au~(\unit[12]{nm})\slash{}\ce{MoO3}~(\unit[50]{nm})) and without capping layer~(\ce{MoO3}~(\unit[12]{nm})\slash{}Au~(\unit[12]{nm})). (b)~Simulated transmittance through complete solar cells with different thickness of the capping layer. The layer sequence was AZO~(\unit[1270]{nm})\slash{}nc-ZnO~(\unit[40]{nm})\slash{}P3HT:PCBM~(\unit[140]{nm})\slash{}\ce{MoO3}~(\unit[12]{nm})\slash{}Au~(\unit[12]{nm})\slash{}\ce{MoO3}~(\unit[$x$]{nm}), with $x$ being the capping layer thickness.}
	\label{fig:fig6}
\end{figure}

As can be seen from Figure~\ref{fig:fig6}a, the application of an additional capping layer~(here exemplarily shown for a thickness of \unit[50]{nm}) significantly increases the transmittance for wavelengths larger than \unit[550]{nm}. Coincidentally, the capping layer gives raise to increased absorption in the UV, resulting in a slightly lower transmittance in that wavelength region. Comparable values for the sheet resistance have been achieved with~(\unit[7.1]{$\Omega$/sq}) and without~(\unit[8.4]{$\Omega$/sq}) the external \ce{MoO3} coating, indicating that the capping layer mainly served as light coupling layer rather than affecting the electrical properties of the electrode.

With the help of TMM simulations, we also studied the influence of the capping layer on the transparency of the whole solar cells. Figure~\ref{fig:fig6}b displays the simulated transmittance through the devices as a function of the wavelength and capping layer thickness. It can be seen that the \ce{MoO3} capping layer enhances the transmittance for wavelengths above the absorption edge of P3HT:PCBM~($>\unit[650]{nm}$). Furthermore, the transmittance is spectrally modulated by the capping layer thickness. The predicted maximum transmittance is about 75--80\%, which is achieved at wavelengths around 660, 730, and \unit[830]{nm}, corresponding to a capping layer thickness of $\sim\!\!50$, $\sim\!\!60$, and \unit[$\sim\!\!75$]{nm}, respectively. Hence, the \ce{MoO3}\slash{}Au\slash{}\ce{MoO3} multilayer anode appears promising for the application in multi-junction devices due to the potentially high transmittance of up to 80\% as well as the possibility to spectrally tune the transmittance by varying the capping layer thickness, the latter being a useful tool in order to match the absorption bands of the underlying subcells.

\begin{figure}
	\centering 
	\includegraphics{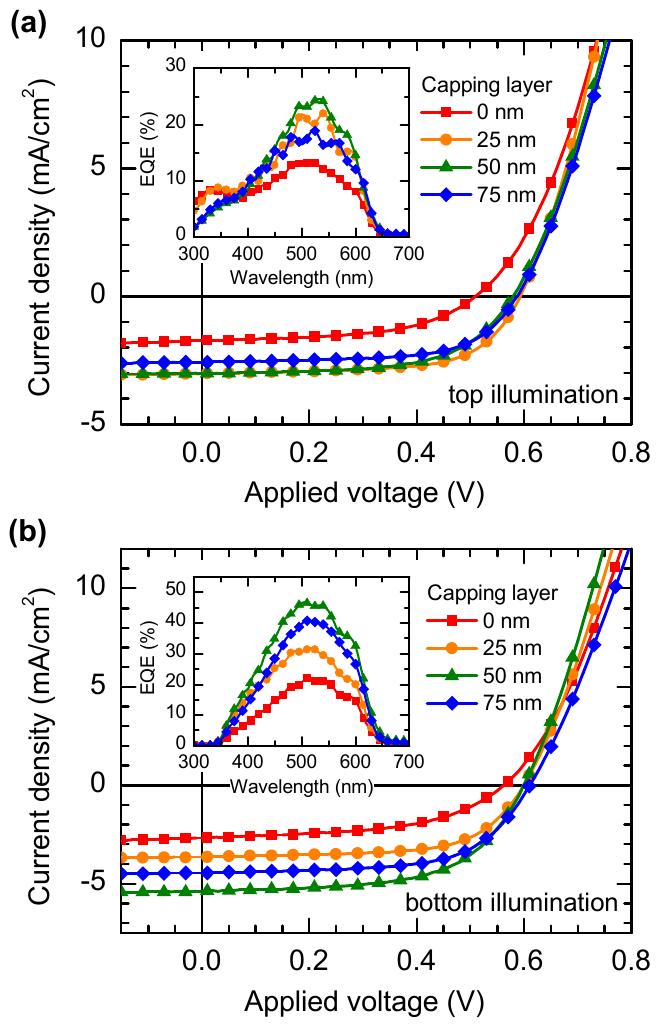}
	\caption{Averaged $J$--$V$ characteristics upon (a)~top and (b)~bottom illumination for semitransparent solar cells with different thickness of the capping layer in the anode system \ce{MoO3}~(\unit[12]{nm})\slash{}Au~(\unit[12]{nm})\slash{}\ce{MoO3}~(\unit[$x$]{nm}) with $x = (0, 25, 50, 75)$. The insets show corresponding EQE spectra for representative devices.}
	\label{fig:fig7}
\end{figure}

To study the effect of the additional capping layer on the electrical performance of the semitransparent solar cells, we built devices with systematically varied capping layer thickness and illuminated them through the bottom and top electrode, respectively. The corresponding $J$--$V$ curves are shown in Figure~\ref{fig:fig7} and statistically evaluated performance data can be found in Table~\ref{tab:tab2}. For both illumination directions, the addition of a capping layer of any thickness had a positive effect on the open-circuit voltage and the fill factor, whereas only slight variations of $V_\text{oc}$ and FF can be seen with the concrete thickness of the capping layer. As the main effect, the capping layer had a strong impact on the short-circuit current density. The solar cells without capping layer showed the smallest values of $J_\text{sc}$ upon both illumination directions. In the case of top illumination, capping layers of \unit[25--50]{nm} thickness increased the short-circuit current density by about 70\%. With thicker \ce{MoO3} layers~(\unit[75]{nm}), $J_\text{sc}$ was found to decrease again, so that an optimum of the capping layer thickness was obtained in the experiment in the range of \unit[25--50]{nm}. In the case of bottom illumination, a qualitatively similar trend was found. However, the variation of the short-circuit current density with the capping layer thickness was much stronger upon bottom illumination with the optimum thickness of \unit[50]{nm} leading to a twofold enhancement of $J_\text{sc}$. Similar trends as in the $J$--$V$ data can also be seen in the EQE curves~(see insets in Figure~\ref{fig:fig7}).

\begin{table*}
\caption{Photovoltaic parameters for semitransparent solar cells with variable capping layer thickness upon top and bottom illumination.\protect\footnote{Cathode: AZO~(\unit[1270]{nm})\slash{}nc-ZnO~(\unit[40]{nm}). Anode: \ce{MoO3}~(\unit[12]{nm})\slash{}Au~(\unit[12]{nm})\slash{}\ce{MoO3}~(\unit[$x$]{nm}) with $x = (0, 25, 50, 75)$. The values reported are mean values averaged over 6--8 individual cells and the corresponding standard deviation.}}
\label{tab:tab2}
\centering
\begin{ruledtabular}
\begin{tabular}{cccccc}
Capping Layer & Illumination &  $V_\text{oc}$~(V) & $J_\text{sc}$~(mA/cm$^2$) & FF & PCE~(\%) \\
\noalign{\vskip .75mm}
\hline
\noalign{\vskip .75mm}
  \unit[0]{nm} & top & 0.51 $\pm$ 0.01	& 1.7	$\pm$ 0.3	& 0.53 $\pm$ 0.02	& 0.5	$\pm$ 0.1 \\
  \unit[25]{nm} & top & 0.59 $\pm$ 0.01	& 3.0	$\pm$ 0.1	& 0.65 $\pm$ 0.01	& 1.2	$\pm$ 0.1 \\
  \unit[50]{nm} & top & 0.58 $\pm$ 0.02	& 3.0	$\pm$ 0.1	& 0.62 $\pm$ 0.03	& 1.1	$\pm$ 0.1 \\
  \unit[75]{nm} & top & 0.59 $\pm$ 0.03	& 2.6	$\pm$ 0.1	& 0.64 $\pm$ 0.01	& 1.0	$\pm$ 0.1 \\[.5em]

  \unit[0]{nm} & bottom & 0.56 $\pm$ 0.03	& 2.7	$\pm$ 0.3	& 0.52 $\pm$ 0.02	& 0.8	$\pm$ 0.2 \\
  \unit[25]{nm} & bottom & 0.60 $\pm$ 0.01	& 3.6	$\pm$ 0.1	& 0.64 $\pm$ 0.01	& 1.4	$\pm$ 0.1 \\
  \unit[50]{nm} & bottom & 0.60 $\pm$ 0.02	& 5.4	$\pm$ 0.2	& 0.60 $\pm$ 0.01	& 1.9	$\pm$ 0.1 \\
  \unit[75]{nm} & bottom & 0.61 $\pm$ 0.01	& 4.4	$\pm$ 0.2	& 0.62 $\pm$ 0.02	& 1.7	$\pm$ 0.1 \\
\end{tabular}
\end{ruledtabular}
\end{table*}

The effect of the capping layer thickness on the photovoltaic performance upon top and bottom illumination has also been studied for semitransparent solar cells with other structurally similar IMI top electrodes, including \ce{MoO3}\slash{}\ce{Ag}\slash{}\ce{MoO3},\cite{Li2014, Tao2009} \ce{MoO3}\slash{}\ce{Ag}\slash{}\ce{V2O5},\cite{Shen2012} and \ce{WO3}\slash{}\ce{Ag}\slash{}\ce{WO3}.\cite{Tao2011} In accordance with our observations, strong variations of the photocurrent density have been reported for both illumination directions, which are typically referred to the altered reflectance\slash{}transmittance properties of the IMI electrodes. However, different functional relations between capping layer thickness and photocurrent have been observed in these studies, ranging from almost linear\cite{Tao2009,Li2014} to oscillating\cite{Tao2011,Shen2012} behavior.

\begin{figure}
	\centering 
	\includegraphics{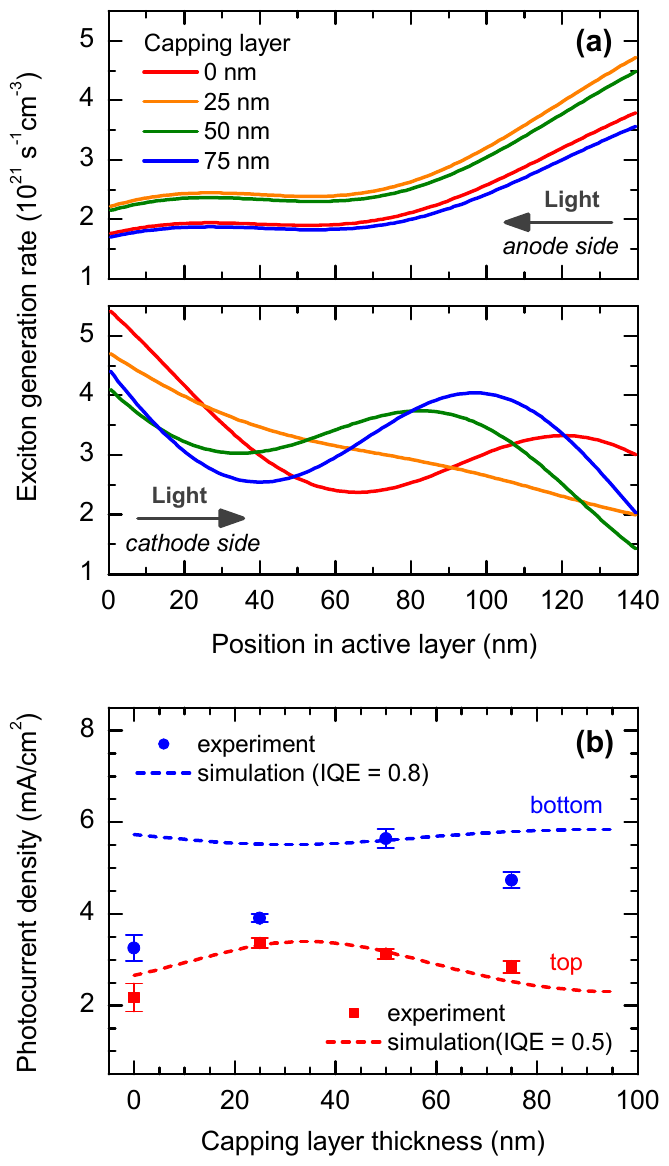}
	\caption{Results of the optical simulations for variable thickness of the \ce{MoO3} capping layer. (a)~Exciton generation rate profiles for AM1.5G illumination either from the anode side~(upper panel) or the cathode side~(lower panel). (b)~Simulated photocurrent density upon top illumination~(red dashed line) and bottom illumination~(blue dashed line), respectively, in comparison to the experimental data under a reverse bias of $\unit[-1]{V}$~(symbols). For the simulated data, internal quantum efficiencies~(IQE) of 0.5~(top illumination) and 0.8~(bottom illumination) were assumed.}
	\label{fig:fig8}
\end{figure}

To corroborate our experimental results, we simulated the distribution of the electric field intensity as a function of the capping layer thickness~(see Figure~S6 in the Supplemental Material\cite{SM}) and calculated the spatially resolved exciton generation rate within the active layer, which is shown in Figure~\ref{fig:fig8}a. In case of illumination through the \ce{MoO3}\slash{}Au\slash{}\ce{MoO3} anode, the exciton generation rate appears relatively unstructured with the maximum generation at the anode\slash{}absorber interface in all cases, and the capping layer thickness affects only the magnitude, but not the shape of the curves. In contrast, upon bottom illumination, pronounced maxima and minima of the exciton generation rate can be seen with their position in the active layer shifting with the capping layer thickness.

Based on the simulated exciton generation rate profiles, we calculated the photocurrent densities to be expected in dependence of the capping layer thickness, and Figure~\ref{fig:fig8}b shows the results in comparison with the experimental data. To account for any voltage-dependent losses, the experimental photocurrents were evaluated under a reverse bias voltage of $\unit[-1]{V}$. However, we note that the values of the photocurrent differed only slightly between reverse bias and short-circuit conditions, which is further underlined by the high fill factors obtained in our experiments. Within the simulations, the IQE was set to constant values of 50\%~(top illumination) and 80\%~(bottom illumination), according to the optimized devices discussed in Section~\ref{sec:overview}. Under these premises, the trend of the experimental data is reasonably reproduced by the optical model in the case of top illumination, in particular the occurrence of the maximum photocurrent for a capping layer thickness of \unit[25--50]{nm}. In the case of bottom illumination, however, the situation is more complicated. Under the assumption of an IQE independent of the capping layer thickness, the optical simulations yielded only slight variations of the photocurrent, which alone could obviously not explain the trend of the experimentally obtained data. Furthermore, the assumption of an IQE of 80\% resulted in a significant overestimation of the photocurrent, except for the optimal capping layer thickness of \unit[50]{nm}. In turn, this indicates that such a high IQE is only achieved for the optimized devices, whereas it is considerably lower in all the other cases.

Additionally, we performed the optical modeling for different thicknesses of the inner \ce{MoO3} layer~(see Supplemental Material,\cite{SM} Figure~S7). This approach would be more related to the optical spacer concept in conventional device architecures.\cite{Kim2006,Gilot2007} Concerning the generation rate profiles, similar trends as for variable capping layer thickness can be seen, whereas the variation of the photocurrent upon top illumination was found to be less pronounced. However, contrary to the capping layer, thickness variations of the internal \ce{MoO3} layer would also affect the charge transport properties of the multilayer electrode and, therefore, were not investigated experimentally in the present work.

If we relate our experimental findings concerning the capping layer thickness to the simulated generation rate profiles~(see Figure~\ref{fig:fig8}a), it appears reasonable to conclude that the IQE depends strongly on the spatial position of the maxima and minima of the exciton generation rate in the active layer. Upon top illumination, the shape of the generation rate profile is independent of the capping layer thickness and the maximum generation rate can be observed at the anode\slash{}absorber interface in all cases. Correspondingly, the trend of the experimental photocurrents can well be described by the simulation under the assumption of a constant IQE. In contrast, the positions of high and low generation rates are strongly dependent on the capping layer thickness in case of bottom illumination, and the assumption of a constant IQE leads to poor agreement between experiment and simulation. However, in the calculation of the current density presented in Figure~\ref{fig:fig8}b, it is assumed that the collection efficiency of separated charge carriers is independent of the spatial position where their precursor excitons are originally generated within the active layer. This is not necessarily true for a real device. Several phenomena can be responsible for a collection efficiency dependent on the spatial exciton generation rate profile in the active layer, three of them will be discussed in more detail in the following.

First, the IQE could be reduced due to the~(non-geminate) recombination of spatially separated charge carriers during charge transport. If this is the case, and comparable mobilities for electrons and hole are assumed, it would be favorable to minimize recombination losses, if the generation rate is maximized in the middle of the active layer, so that electrons and holes will on average have balanced distances to travel to the electrodes. According to Figure~\ref{fig:fig8}a, such a situation would be realized for bottom illumination with an intermediate capping layer thickness of \unit[50]{nm}. For this thickness, the generation rate is maximized in the middle of the active layer and minimized at the borders. This coincides with the maximum photocurrent observed experimentally and indicates a high collection efficiency for separated charge carriers in this case, which could be related to a low recombination rate. In contrast, without a capping layer, the optical simulations revealed high exciton generation rates at the borders and a low generation rate in the middle of the active layer. Correspondingly, the lowest current density was found in the experiment, indicating a relatively low collection efficiency, possibly related to a high recombination rate. However, from the electrical behavior of our solar cells under illumination, the occurrence of significant bimolecular recombination losses appears unlikely, as one would expect a stronger voltage-dependency of the photocurrent and, correspondingly, reduced values of the fill factor in that case.\cite{Cowan2010,Cowan2013} Hence, if non-geminate recombination is the dominant mechanism here, we suppose it to be mainly of monomolecular nature.

Another possible explanation is given by Moul\'{e} and Meerholz,\cite{Moule2008} who suggested a reduced probability for exciton separation~(i.e., increased exciton recombination) in the vicinity of the electrode interfaces compared to the middle of the active layer. Consequently, the separation of photogenerated excitons would become less efficient  when the exciton generation rate is high at the boundaries of the active layer, as has been experimentally demonstrated in that study for the interface between poly (3,4-ethylenedioxythiophene):poly (styrenesulfonate)~(PEDOT:PSS) and P3TH:PCBM. Although it is unclear whether this finding could be generalized for interfacial materials other than PEDOT:PSS, the proposed exciton recombination close to electrodes could provide a further hint, why the exciton generation profile upon top illumination with a high generation rate at the anode\slash{}absorber interface appears principally unfavorable compared to the case of bottom illumination.

Finally, the variation of the IQE could also stem from vertical composition profiles within the polymer--fullerene blend layer. It has been found in numerous studies that the nanomorphology of the binary polymer--fullerene blends in the active layer of organic BHJ solar cells can strongly depend on preparation parameters like the blend composition, the regio-regularity of the polymer, or applied annealing conditions.\cite{Guo2013,Kohn2013,vanBavel2009} Van Bavel et al.\cite{vanBavel2009} showed by means of electron tomography that the volume fractions of P3HT and PCBM are not necessarily equally distributed throughout the P3HT:PCBM blend layer, which is commonly neglected in optical simulations. For instance, the authors observed a lower percentage of P3HT close to the top of spin-coated films compared to the bottom. As P3HT is the main absorbing part of the blend layer, the latter situation would lead to an overestimation of the photocurrents in case of high electric field intensities close to the upper boundary of the active layer. For our devices, this would be the case for top illumination and, indeed, may explain the lower IQE values observed for that illumination direction. However, we note that the concrete vertical composition profile strongly depends on the preparation conditions of the active layer.

Thus, to conclude on this section, electrical losses, either during exciton separation or charge transport, as well as vertical composition gradients within the active layer are suggested as possible reasons for the strong variation of the photocurrents (i)~between top and bottom illumination and (ii)~with the capping layer thickness upon bottom illumination. The optical simulations provided a reasonable understanding of these findings in terms of the spatial distribution of the exciton generation rate. However, to finally judge which of the proposed effects play the dominating role in our case would require further investigations and is beyond the scope of the present study.

\subsection{Variation of the Active Layer Thickness}
\label{sec:var-thickness}

\begin{figure}
	\centering 
	\includegraphics{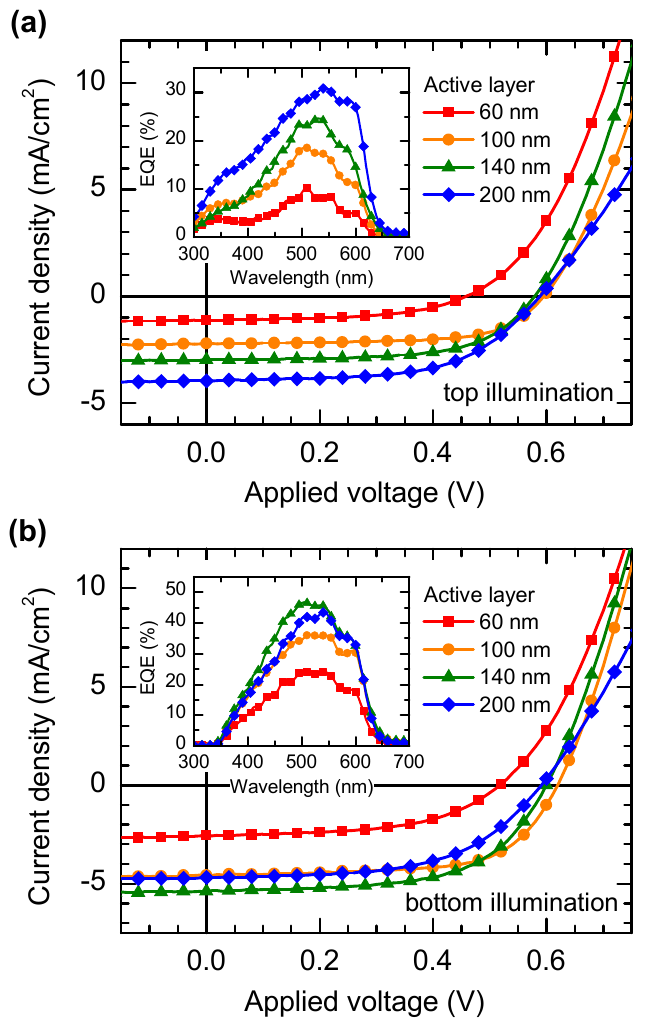}
	\caption{
Averaged $J$--$V$ characteristics upon (a)~top and (b)~bottom illumination for semitransparent solar cells with the optimized contact configuration and different active layer thicknesses~(60, 100, 140, and \unit[200]{nm}). The insets show corresponding EQE spectra for representative devices.}
	\label{fig:fig9}
\end{figure}
 
One of the most crucial issues for photocurrent generation is the thickness of the active layer. On the one hand, the active layer must be thick enough to absorb enough light in order to achieve efficient light harvesting. On the other hand, too thick active layers may lead to increased electrical losses during charge transport because of the limited charge carrier mobility. In the present work, we prepared a series of semitransparent solar cells with the optimized electrode configuration and different active layer thicknesses ranging from \unit[60]{nm} to \unit[200]{nm}. The thickness variation has been obtained by adjusting the concentration of the P3HT:PCBM solutions used for spin-coating. Figure~\ref{fig:fig9} and Table~\ref{tab:tab3} show and summarize performance data obtained from $J$--$V$ measurements. It can be seen that the lowest active layer thickness~(\unit[60]{nm}) resulted in poor device performance with PCEs below 1\% and relatively small values of the open-circuit voltage and fill factor. In case of the thicker active layers, the values of $V_\text{oc}$ and FF are significantly improved. However, at a closer look, a tendency can be observed regarding the fill factor, which reaches a maximum at a medium thickness of \unit[100]{nm} upon both illumination directions, but is found to decrease again if the active layer thickness is further increased. Concerning the obtained values for the photocurrent density, a clear trend can be observed upon illumination through the anode: the thicker the active layer, the higher the photocurrent. In contrast, in the case of bottom illumination, the current density increases with the absorber layer thickness up to \unit[140]{nm}, but then decreases again for the thickest active layers prepared~(\unit[200]{nm}).

\begin{table*}
\caption{Photovoltaic parameters for semitransparent solar cells with variable thickness of the active layer upon top and bottom illumination.\protect\footnote{Cathode: AZO~(\unit[1270]{nm})\slash{}nc-ZnO~(\unit[40]{nm}). Anode: \ce{MoO3}~(\unit[12]{nm})\slash{}Au~(\unit[12]{nm})\slash{}\ce{MoO3}~(\unit[50]{nm}). The values reported are mean values averaged over 6-8 individual cells and the corresponding standard deviation.}}
\label{tab:tab3}
\centering
\begin{ruledtabular}
\begin{tabular}{cccccc}
Active layer & Illumination &  $V_\text{oc}$~(V) & $J_\text{sc}$~(mA/cm$^2$) & FF & PCE~(\%) \\
\noalign{\vskip .75mm}
\hline
\noalign{\vskip .75mm} 
  \unit[60]{nm} & top & 0.46 $\pm$ 0.05	& 1.1 $\pm$ 0.3	&	0.53 $\pm$ 0.05	&	0.3 $\pm$ 0.1 \\
  \unit[100]{nm} & top & 0.60 $\pm$ 0.01	&	2.2	$\pm$ 0.1	&	0.66 $\pm$ 0.01	&	0.9 $\pm$ 0.1 \\
  \unit[140]{nm} & top & 0.58 $\pm$ 0.02	&	3.0	$\pm$ 0.1	&	0.62 $\pm$ 0.03	&	1.1 $\pm$ 0.1 \\
  \unit[200]{nm} & top & 0.59 $\pm$ 0.01	&	4.0	$\pm$ 0.4	&	0.58 $\pm$ 0.01	&	1.4 $\pm$ 0.1 \\[.5em]

  \unit[60]{nm} & bottom & 0.52 $\pm$ 0.06	&	2.6 $\pm$ 0.6	&	0.55 $\pm$ 0.05	&	0.8 $\pm$ 0.3 \\
  \unit[100]{nm} & bottom & 0.62 $\pm$ 0.01	&	4.6 $\pm$ 0.3	&	0.65 $\pm$ 0.01	&	1.9 $\pm$ 0.1 \\
  \unit[140]{nm} & bottom & 0.60 $\pm$ 0.02	&	5.4 $\pm$ 0.2	&	0.60 $\pm$ 0.01	&	1.9 $\pm$ 0.1 \\
  \unit[200]{nm} & bottom & 0.59 $\pm$ 0.01	&	4.7 $\pm$ 0.3	&	0.56 $\pm$ 0.03 & 1.6 $\pm$ 0.1 \\
\end{tabular}
\end{ruledtabular}
\end{table*}

\begin{figure}
	\centering 
	\includegraphics{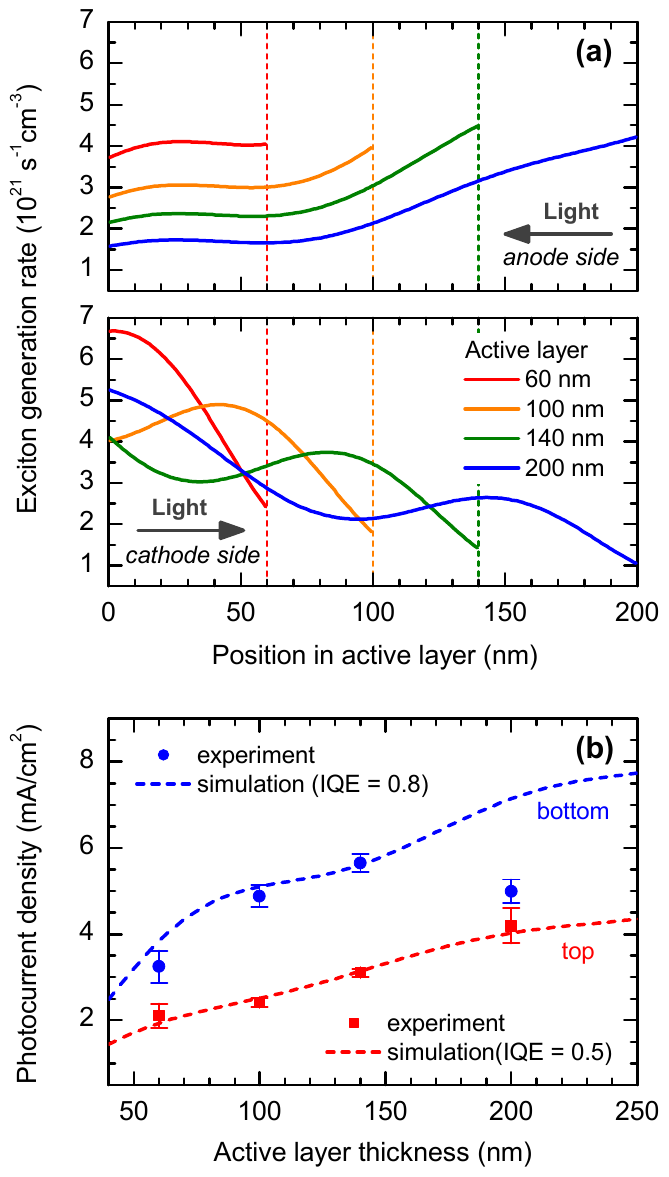}
	\caption{
Results of the optical simulations for variable thickness of the active layer. (a)~Exciton generation rate profiles for AM1.5G illumination either from the the anode side~(upper panel) or the cathode side~(lower panel). (b)~Simulated photocurrent density upon top illumination~(red dashed line) and bottom illumination~(blue dashed line), respectively, in comparison to the experimental data under a reverse bias of $\unit[-1]{V}$~(symbols). For the simulated data, internal quantum efficiencies~(IQE) of 0.5~(top illumination) and 0.8~(bottom illumination) were assumed.}
	\label{fig:fig10}
\end{figure}

To investigate, whether the different behavior can be explained by optical effects, we performed TMM simulations. Figure~\ref{fig:fig10}a shows the resulting exciton generation rate for the different active layer thicknesses, and Figure~\ref{fig:fig10}b displays the simulated photocurrent density~(dashed lines) in comparison to the experimental data obtained under reverse bias~(symbols). Again, we assumed constant IQEs of 50\%~(top illumination) and 80\%~(bottom illumination). For both illumination directions, the simulated photocurrent exhibits a monotonous increase with the absorber layer thickness, being more structured upon bottom illumination due to the more pronounced interference effects when the device is illuminated through the transparent cathode. A reasonable agreement can be seen between the simulated and measured photocurrents upon top illumination, indicating the IQE to be independent of the active layer thickness in this case. Upon bottom illumination, we also observed a good match between simulation and experiment for low and medium thicknesses of the active layer. However, in case of the largest thickness under consideration~(\unit[200]{nm}), significantly lower photocurrents than expected by the simulation were obtained. This indicates that the IQE is decreasing for larger active layer thicknesses when the device is illuminated through the bottom electrode, as it has also been observed by other authors in the case of non-transparent polymer solar cells,\cite{Gilot2007,Armin2014,Chen2014} and is commonly attributed to electrical losses.

In principle, it is well known that charge transport can become a limiting factor in organic solar cells, if the absorber layer becomes too thick. The question arises, however, why this should lead to a reduced photocurrent only under bottom illumination. Here, the optical simulations reveal a significant difference: If the devices are illuminated through the anode, the exciton generation rate profile has a pronounced maximum close to the anode and then decreases inside the absorber layer~(see Figure~\ref{fig:fig10}a). In contrast, if illuminated through the AZO electrode, the generation rate is higher on the cathode side than on the anode side of the absorber layer. This difference has an impact on the mean distance which electrons, respectively holes, have to travel from the point of charge carrier generation to the selective electrodes. For top illumination, the generation rate profile appears favorable for hole transport, whereas the distance for electrons generated near the anode towards the cathode is relatively long. Vice versa, holes have on average a long distance to travel towards the anode in the case of bottom illumination. If hole transport would be more severely limiting than electron transport, this might provide an explanation why the photocurrent shows a different behavior for the two illumination directions in the experiments. 

The statement that the performance is mainly limited by charge transport in case of thicker active layers is also supported by the decreasing values of FF with increasing active layer thickness. This points to electrical losses around the maximum power point for thick active layers and may be reasonably connected to non-geminate recombination, e.g., of bimolecular nature. However, it is noteworthy that vertical phase separation could also be of importance here, which would make the situation more complex. Detailed studies of the charge transport through the active layer might provide deeper insight, but would go beyond the scope of the present work.  Thus, the optical simulations cannot directly explain the experimentally found maximum of the photocurrent for devices with \unit[140]{nm} thick absorber layers under bottom illumination, but provide at least a hint that the effect may be related to the nature of the exciton generation rate profiles.

An important issue related to polymer solar cells is the light-induced degradation of the active layer. During continuous operation, significant alterations of the active layer morphology have been recognized,\cite{Schaffer2013,Manceau2011} resulting in deteriorated photovoltaic performance. Another well-known degradation pathway is the photo-oxidation of the polymer phase in the presence of oxygen, which has been found to exhibit a strong dependence on the excitation wavelength.\cite{Hintz2012} Although investigations of the long-term stability were not in the focus of the present work, we believe that semitransparent solar cells with adjustable optical properties can be utilized in order to answer the question of how these degradation mechanisms are affected by the spatial exciton generation rate profile throughout the active layer. This could provide deep insight into the possible depth-dependence of the active layer degradation and is an interesting topic for future research.


\subsection{Summary and Conclusion}
In summary, we have investigated the optimization of the electron and hole contact in ITO-free semitransparent organic solar cells. Concerning the cathode, sputter-deposited AZO electrodes were demonstrated to be comparably suitable as ITO in combination with an additional layer of (intrinsic) nc-ZnO. At the anode side, \ce{MoO3}\slash{}Au\slash{}\ce{MoO3} multilayer systems with sheet resistances lower than \unit[9]{$\Omega$/sq} could be produced and successfully applied to the semitransparent solar cells. The additional \ce{MoO3} capping layer was found to enhance light coupling and increase the transparency of the devices in the spectral range above \unit[650]{nm}. Furthermore, the capping layer increased the performance of the solar cells, mainly related to an improvement of the photocurrent. Systematic variations of the thickness of the capping layer and the active layer were performed, and the device performance was investigated as dependent on the illumination direction. Finally, we can report on a power conversion efficiency of 2.0\% for optimized semitransparent ITO-free organic solar cells based on P3HT:PCBM with a maximum transmittance of 60\% for wavelength above \unit[650]{nm}. 

The experimental work was accompanied by optical simulations of the optical electric field intensity distribution within the semitransparent solar cells. From the field intensity distribution, the overall transmittance of the solar cells could be calculated as well as spatially resolved profiles of the exciton generation rate and, finally, the expectable photocurrent densities. In some cases, the simulations provided direct explanations for the experimentally observed current variations. For example, the photocurrent density showed a pronounced maximum for capping layers thicknesses of about \unit[25--50]{nm} if the devices were illuminated through the anode. This maximum could be explained by the optical interference effects. In other cases, differences were observed between experimental and calculated photocurrent densities, which indicates that the device performance is not governed by optical effects alone. Nevertheless, the simulations turned out to be still a useful tool, because the generation rate profiles provided a theoretical basis for the development of interpretations for the experimentally observed dependences of the photocurrent density on layer thickness parameters, especially concerning the recombination of photogenerated charge carriers.

Finally, from a more general point of view, we would like to emphasize that the electrode concepts as well as the simulation approach are expected to be transferable also to organic solar cells with other absorber systems than P3HT:PCBM.


\begin{acknowledgements}
The authors thank Kambulakwao Chakanga and Omid Madani Ghahfarokhi~(Next Energy) for the preparation of AZO films, Janet Neerken and Matthias Macke~(University of Oldenburg) for practical support, and Michael Richter~(University of Oldenburg) for assistance with preliminary work on the optical simulations. Financial support from the EWE-Nachwuchsgruppe ``D\"unnschichtphotovoltaik'' by the EWE AG, Oldenburg, Germany is gratefully acknowledged.
\end{acknowledgements}

\bibliography{references}

\end{document}


\title{Supplemental Material for ``Semitransparent Polymer-Based Solar Cells with Aluminum-Doped Zinc Oxide Electrodes''}

\author{Sebastian Wilken}
\email{sebastian.wilken@uol.de}
\author{Verena Wilkens}
\author{Dorothea Scheunemann}
\affiliation{Institute of Physics, Energy and Semiconductor Research Laboratory, Carl von Ossietzky University of Oldenburg, 26111 Oldenburg, Germany}
\author{Regina-Elisabeth Nowak}
\author{Karsten von Maydell}
\affiliation{NEXT ENERGY, EWE Research Centre for Energy Technology, Carl-von-Ossietzky-Str.~15, 26129 Oldenburg, Germany}
\author{J\"urgen Parisi}
\author{Holger Borchert}
\affiliation{Institute of Physics, Energy and Semiconductor Research Laboratory, Carl von Ossietzky University of Oldenburg, 26111 Oldenburg, Germany}

\maketitle
\tableofcontents

\section{Transmittance and resistance of AZO films}
\begin{figure}[h!]
\includegraphics[width=0.6\textwidth]{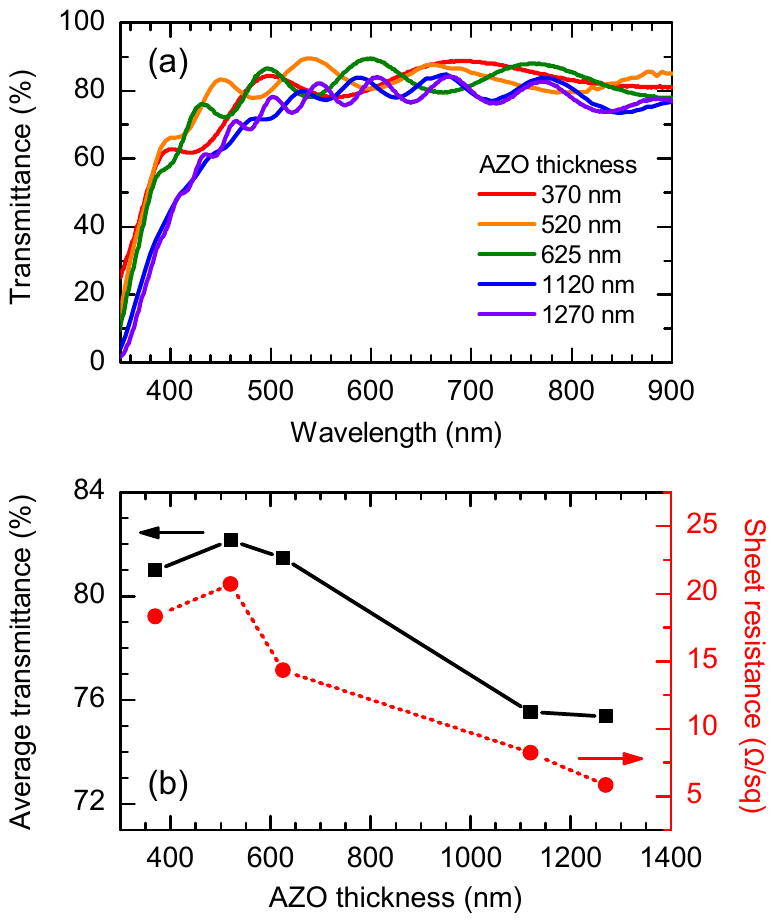}%
\caption{(a) Spectral transmittance of AZO layers with different thicknesses ranging from \unit[370]{nm} to \unit[1270]{nm}. (b) Average transmittance (\unit[400--900]{nm}) and sheet resistance of the AZO layers versus film thickness.}
\label{fig:AZO}
\end{figure}

Figure \ref{fig:AZO}a shows transmittance spectra of AZO layers, sputtered on glass substrates, with different thicknesses (\unit[370]{nm}, \unit[520]{nm}, \unit[625]{nm}, \unit[1120]{nm}, and \unit[1270]{nm}). In Figure \ref{fig:AZO}b, the average transmittance for wavelength between 400 and \unit[900]{nm} is plotted against the layer thickness together with the sheet resistance, obtained from a linear four point probe measurement. Therefore, four equally spaced gold probes were used to contact the samples and the resistance was measured with a source measurement unit (Keithley 2400).

\clearpage
\section{Determination of optical constants}
\begin{figure}[h!]
\includegraphics{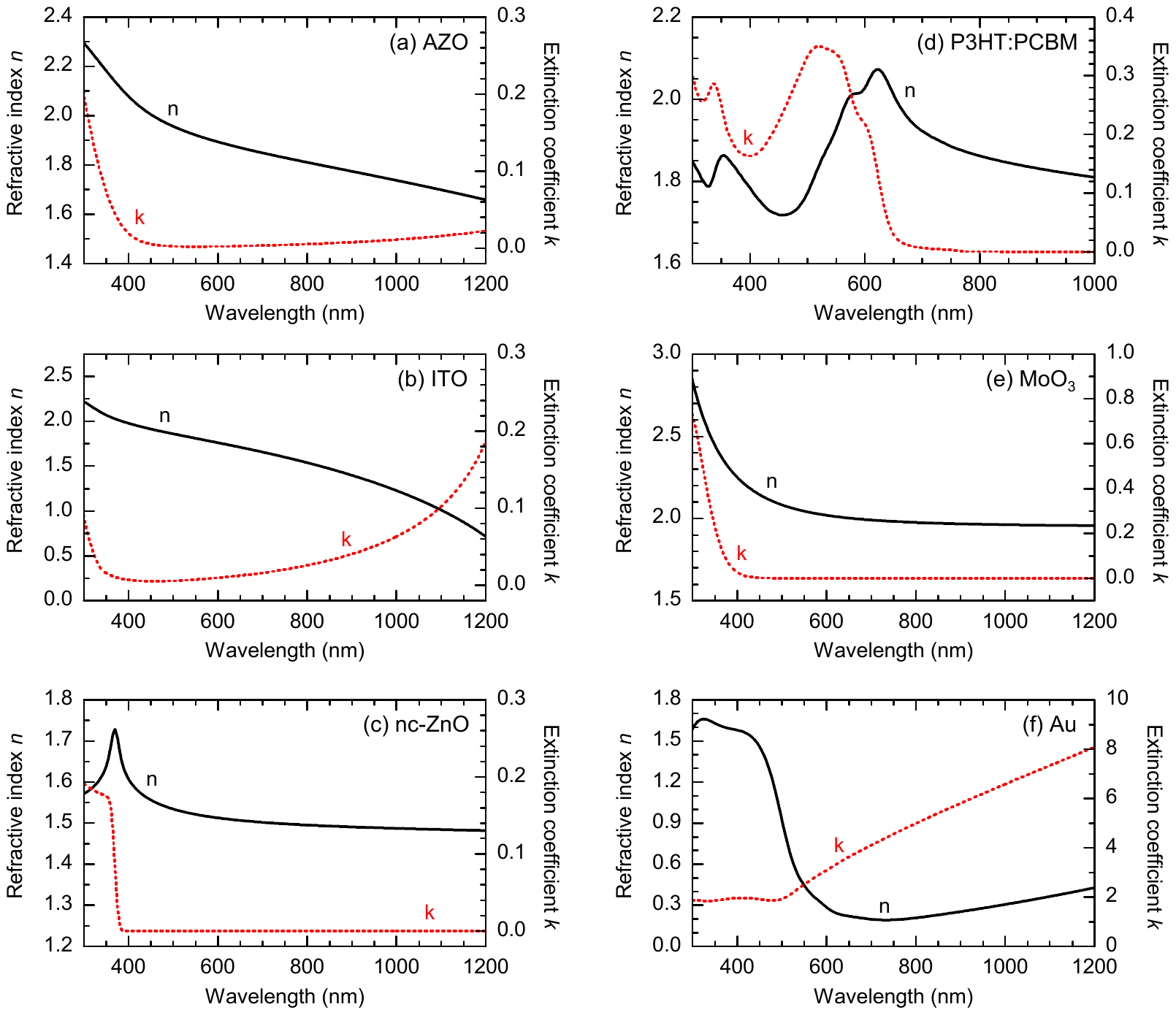}%
\caption{Optical constants determined by means of modeling experimentally obtained SE data for (a) AZO, (b) ITO, (c) nc-ZnO, (d) P3HT:PCBM, (e) \ce{MoO3}, and (f) Au.}
\label{fig:nk}
\end{figure}

The optical constants of the materials used for solar cell fabrication, i.e., the refractive index $n$ and the extinction coefficient $k$, were obtained via modeling of spectroscopic ellipsometry (SE) data. The results for AZO, ITO, nc-ZnO, P3HT:PCBM, \ce{MoO3}, and Au are summarized in Figure \ref{fig:nk}. 

SE measurements were performed on a rotating analyzer ellipsometer (J. A. Woollam VASE) for wavelengths between 280 and \unit[1700]{nm}. The samples were prepared either on glass substrates or Si wafers with a \unit[2]{nm} native oxide layer on top. The optical constants of the substrates were determined prior to the measurements. Depending on the substrate, ellipsometric data ($\Phi,\Delta$) was obtained under three angles of incidence, 55$^\circ$, 65$^\circ$, and 75$^\circ$ (glass) and 65$^\circ$, 70$^\circ$, and 75$^\circ$ (Si), respectively. The experimental $\Phi,\Delta$ data was analyzed with the software \texttt{WVASE32} (version 3.768). In the following, we present some details regarding the sample preparation and the modeling of the data. Herein, the mean squared error (MSE) is used to indicate how well the model fits the data.

\subsection{Aluminum-doped zinc oxide (AZO)}
Transmission and SE measurements of an AZO film on glass with \unit[324]{nm} thickness were fitted simultaneously to increase the quality of the model fit. The dielectric properties of AZO were modeled with the OJL~model\cite{OLeary1997} to describe the energy range near the band gap and the extended Drude model to account for free carrier absorption of the electrons.\cite{Mergel2002} A Bruggemann effective medium layer~(EMA) on top of the bulk AZO layer was added to model the surface roughness.

\subsection{Indium tin oxide (ITO)}
The ITO covered glass substrates were commercially purchased (Pr{\"a}zisions Glas \& Optik GmbH) with a thin passivation layer of \ce{SiO2} located between the ITO film and the glass substrate. In detail, the layer sequence was glass (\unit[1.1]{mm})/\ce{SiO2} (\unit[25]{nm})/ITO (\unit[200]{nm}). First, we acquired SE data of a glass/\ce{SiO2} substrate obtained by etching away the ITO film with hydrochloric acid, and modeled the optical constant using Cauchy models. Then, the ($\Phi,\Delta$) data of a complete glass/\ce{SiO2}/ITO sample was recorded and analyzed with the as-obtained optical constants of the substrate serving as input parameters. To describe the ITO layer, we employed a generalized oscillator model consisting of one Tauc-Lorentz, one rho-tau Drude, and one Gaussian oscillator. Additionally, a Bruggeman effective medium approximation~(EMA) layer~(50\% void) with a thickness of \unit[2.5]{nm} was introduced on top of the ITO layer to account for the surface roughness. Finally, we obtained a MSE of 6.4.

\subsection{Zinc oxide nanocrystals (nc-ZnO)}
SE measurements were performed on a spin-coated film of ZnO nanoparticles (thickness \unit[72]{nm}) on Si. Experimental data was modeled with a generalized oscillator model consisting of a Herzinger--Johs parametric oscillator model function~(Psemi-M0) and two Gaussian oscillators with a MSE of 3.6. Additionally, we observed high correlation between the $k$ values obtained via the modeling of SE data and the spectral absorbance measured on colloidal ZnO solutions with a UV/Vis spectrophotometer.

\subsection{Active layer (P3HT:PCBM)}
For the investigations on P3HT:PCBM blend layers, we prepared samples via spin-coating either on Si or glass substrates. SE data was collected for the Si-based samples, and the films on glass served for transmittance measurements. To account for the fact that the blend layer is a mixture of two materials, we used a Bruggeman effective medium approximation~(EMA) layer to model the data, consisting to $\sim\!\!50\%$ of each P3HT and PCBM. For the P3HT component of the EMA layer, we used a generalized oscillator model consisting of seven single Gaussian oscillators to describe the different absorption features of P3HT. For the PCBM component, we used data provided free of charge by the group of Michael McGehee~(Stanford University).\cite{Burkhard2010} The model was mainly forced to fit the transmittance data and the finally obtained MSE was $\sim\!\!1$. We also cross-checked the model with the SE data, which yielded larger MSE values of 25--30. However, the model is believed to be acceptable for our purpose. This is also supported by the work of Burkhard et al.,\cite{Burkhard2010} who found that even under the simple assumption of a constant value for the refractive index of $n = 2$, the active absorption could accurately be modeled using transfer matrix simulations. Their results showed only negligible discrepancies compared to the case where the ``real'' optical constants (obtained from SE measurements) were used.

\subsection{Molybdenum oxide (\ce{MoO3})}
SE measurements were performed on a thermally evaporated thin film on glass with a thickness of \unit[48]{nm}. The index of refraction $n$ was modeled using the three-coefficient Cauchy model. For the extinction coefficient $k$, an approach similar to the nc-ZnO layers was used to account for the absorption in the UV. Finally, we obtained an MSE of 3.5.

\subsection{Gold (\ce{Au})}
An ultra-thin Au film with a thickness of \unit[14.5]{nm}, thermally evaporated on a glass substrate, was used for the SE measurements. Experimental data was modeled with a generalized oscillator model consisting of two Tauc-Lorentz, three Gaussian, and one Lorentz oscillator. A very low MSE of 0.8 has been obtained after fitting, indicating a high accuracy of the model.

\clearpage
\section{Evaluation of series resistance}
\begin{figure}[h!]
\includegraphics[width=0.6\textwidth]{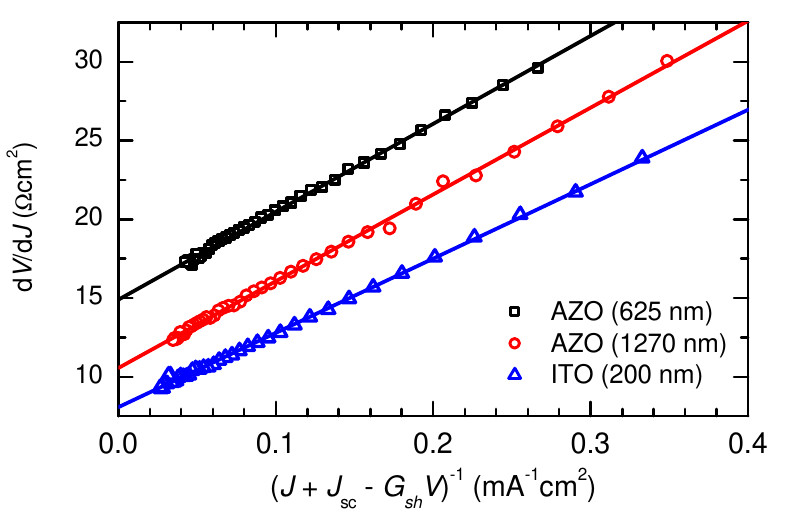}%
\caption{Evaluation of the series resistance for exemplary semitransparent solar cells with different cathode materials, AZO (thickness \unit[625]{nm}), AZO (\unit[1270]{nm}), and ITO (\unit[200]{nm}). The corresponding $J$--$V$ curves where recorded upon illumination from the cathode side.}
\label{fig:Rs}
\end{figure}

The series resistance was derived from the $J$--$V$ characteristics using a method introduced by Hegedus and Shafarman.\cite{Hegedus2004} Hence, the derivative $\text{d}V/\text{d}J$ is plotted against the inverse current density $(J+J_\text{sc}-G_\text{sh}V)^{-1}$. To account for the additional photocurrent under illumination, the current density was corrected by $J_\text{sc}$. Further correction has been made by the non-zero shunt conductance $G_\text{sh}$ (i.e., a finite parallel resistance $R_\text{p}$), which effectively quantifies the steepness of the $J$--$V$ curve around $J_\text{sc}$. The series resistance is then given by
\begin{equation}
\frac{\text{d}V}{\text{d}J} = R_\text{s} + \frac{AkT}{q} (J + J_\text{sc} - G_\text{sh}V)^{-1},
\label{eq.Rs}
\end{equation}
where $A$ is the diode ideality factor, $k$ Boltzmann's constant, $T$ the absolute temperature, and $q$ the elementary charge. A linear fit of the data gives an intercept of $R_\text{S}$ and a slope $AkT/q$. In Figure \ref{fig:Rs}, we show exemplary plots according to Eq.~(\ref{eq.Rs}) for semitransparent solar cells with different cathode materials (AZO with two different thicknesses vs. ITO).

\clearpage
\section{Transmittance and resistance of ultra-thin metal films}
\begin{figure}[h!]
\includegraphics[width=0.6\textwidth]{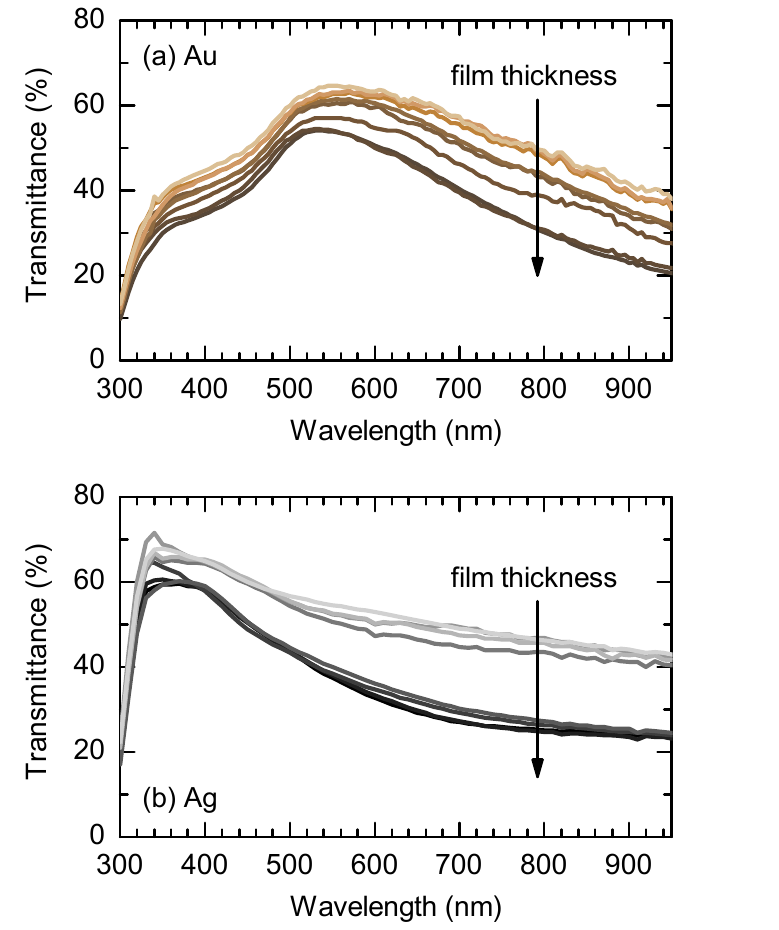}%
\caption{Spectral transmittance of ultra-thin (a) Au and (b) Ag films with varying film thickness.}
\label{fig:Au_Ag_Trans}
\end{figure}

Figure \ref{fig:Au_Ag_Trans} shows the spectral transmittance of ultra-thin metal films, made from Au and Ag. The films were thermally evaporated on glass substrates covered with a \unit[12]{nm} thick \ce{MoO3} layer. The nominal thickness of the films, estimated in situ with a calibrated quartz microbalance, ranged from \unit[8]{nm} to \unit[15]{nm}.

The sheet resistance of the ultra-thin metal films was determined using the van der Pauw method\cite{vanderPauw1958}. Therefore, the samples were electrically contacted with four microprobes placed in the corners of the rectangular thin films (here named 1, 2, 3, and 4 in a clockwise or counterclockwise order). Using a source measurement unit (Keithley 2400), a current flow is forced along one of the edges of the films, say between contacts 1 and 2 ($I_{12}$) and, simultaneously, the voltage drop along the opposite edge is measured ($V_{34}$). The resistance across the area delimited by the two pairs of contacts is calculated according to Ohm's law (in this example, $R_{12,34} = V_{34}/I_{12}$). The contacts are then circularly shifted until all possible (four) contact configurations have been reached and, afterwards, the average resistance across the two horizontal and two vertical contact configurations was calculated ($R_\text{horiz.}$, $R_\text{vert.}$). Because of the rectangular shape of the films, $R_\text{horiz.}$ and $R_\text{vert.}$ differed by a factor of $\sim\!\!4$, which was respected by using the correction factor $f$ introduced in the original publication of van der Pauw\cite{vanderPauw1958}. Finally, the sheet resistance was calculated via
\begin{equation}
R_\text{sheet} = \frac{\pi}{\ln 2} \frac{R_\text{horiz.} + R_\text{vert.}}{2} \cdot f.
\end{equation}

\begin{figure}
\includegraphics[width=0.85\textwidth]{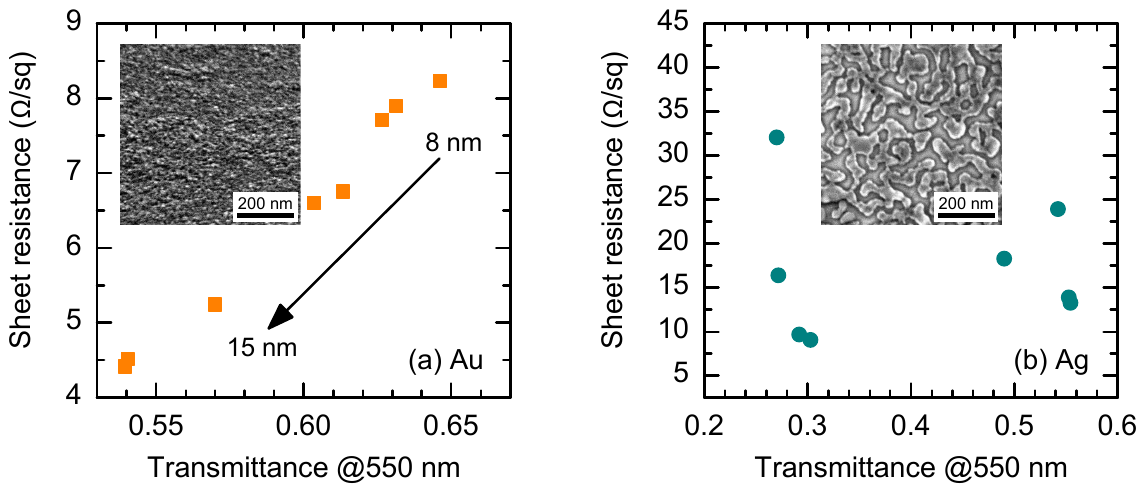}%
\caption{Sheet resistance versus optical transmittance at a wavelength of \unit[550]{nm} for ultra-thin films of (a) Au and (b) Ag, respectively. The films were thermally evaporated on glass substrates, covered with a \unit[12]{nm} thick \ce{MoO3} layer. The insets show representative SEM images of the as-obtained metal films.}
\label{fig:Au_Ag}
\end{figure}

In Figure \ref{fig:Au_Ag}, the sheet resistance of the Au and Ag films is plotted against the transmittance at a wavelength of \unit[550]{nm}. We note that we used the transmittance on the abscissa instead of the film thickness, because common measurements techniques for the film thickness are supposed to be too error-prone in that range. As to be expected, a straight relation between transmittance and sheet resistance can be seen in case of Au. The SEM image (inset) visualizes the homogeneity of the Au films. In contrast, no correlation between transmittance and sheet resistance can be seen in case of ultra-thin Ag films, as it is expected for film thicknesses below the percolation threshold. This is further underlined by the SEM images, showing incomplete surface coverage with isolated Ag islands rather than a closed layer.

\clearpage
\section{Field distribution for different capping layer thicknesses}
\begin{figure}[h!]
\includegraphics[width=\textwidth]{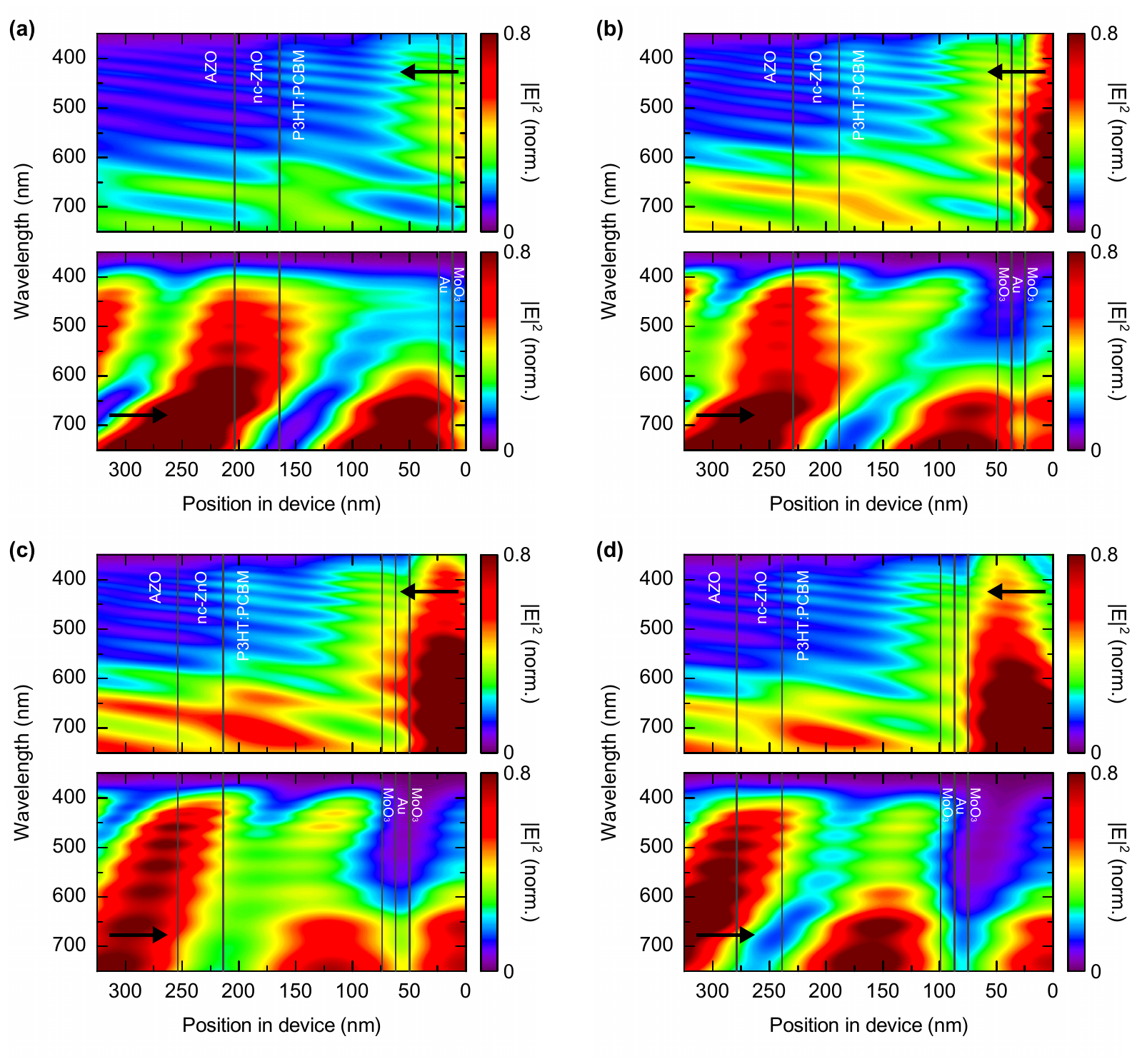}%
\caption{Simulated electric field intensity $|E|^2$ within the semitransparent solar cells with a capping layer thickness of (a) \unit[0]{nm}, (b) \unit[25]{nm}, (c) \unit[50]{nm}, and (d) \unit[75]{nm} upon illumination via the \ce{MoO3}/Au/\ce{MoO3} anode (upper panels) and the AZO/nc-ZnO cathode (lower panels), respectively. Arrows indicate the direction of incident light.}
\label{fig:fig_S6}
\end{figure}

\clearpage
\section{Variation of inner \ce{MoO3} layer thickness}
\begin{figure}[h!]
\includegraphics[width=0.5\textwidth]{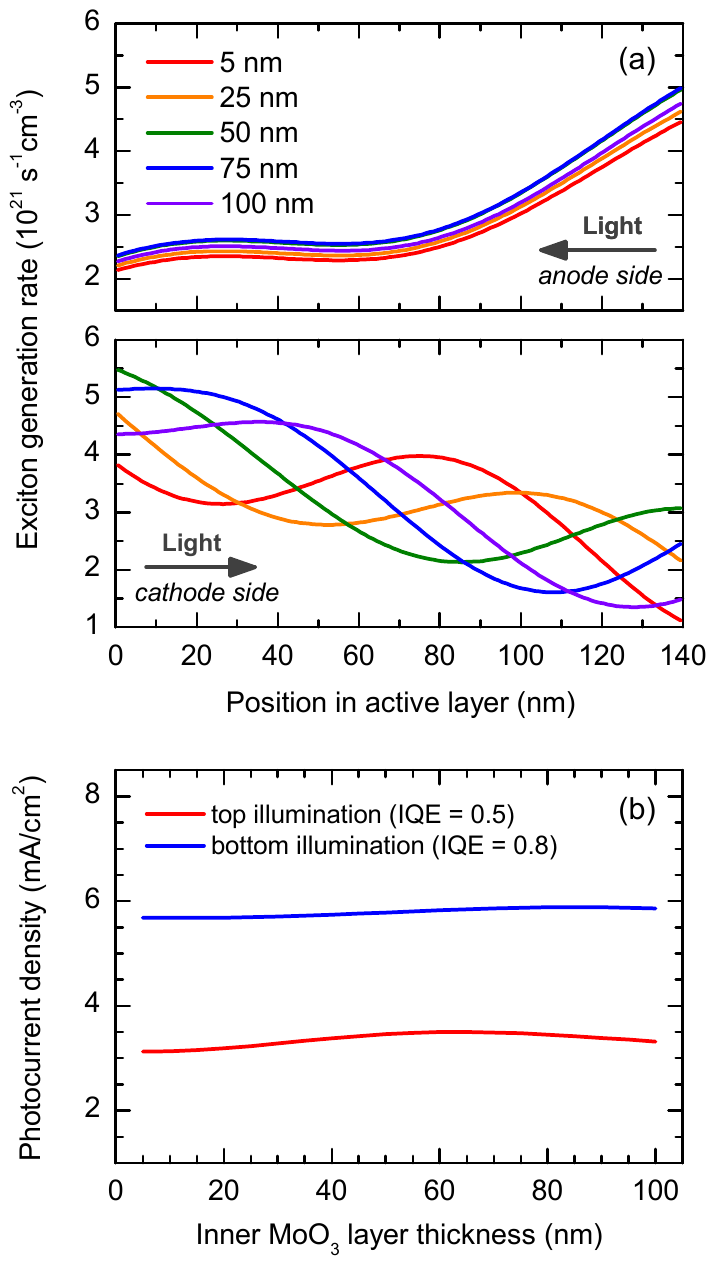}%
\caption{Simulated exciton generation rate (a) and photocurrent density (b) for varying thickness of the inner \ce{MoO3} layer in the \ce{MoO3}/Au/\ce{MoO3} multilayer electrode upon top and bottom illumination, respectively. The thickness of the capping layer was fixed to \unit[50]{nm}. For the calculation of the photocurrents, internal quantum efficiencies (IQE) of 0.5 (top illumination) and 0.8 (bottom illumination) were assumed.}
\label{fig:fig_S7}
\end{figure}

Figure \ref{fig:fig_S7} shows the simulated exciton generation rates and photocurrents for various thicknesses of the inner \ce{MoO3} layer in the \ce{MoO3}/Au/\ce{MoO3} multilayer electrode. In this simulations, the thickness of the outer \ce{MoO3} capping layer was set to a fixed value of \unit[50]{nm}.

\bibliography{references}